\documentclass{raa}
\usepackage{graphicx,times}             
\usepackage{natbib}
\usepackage{amssymb,amsmath}
\usepackage{longtable}
\usepackage[a4paper=true]{hyperref}

\begin{document}

  \title{The Binding Energy Parameter for Common Envelope Evolution
}

   \volnopage{Vol.0 (200x) No.0, 000--000}      
   \setcounter{page}{1}          

   \author{Chen Wang
      \inst{1,2}
   \and Kun Jia
      \inst{1,2}
   \and Xiang-Dong Li
      \inst{1,2}
   }

   \institute{Department of Astronomy, Nanjing University, Nanjing 210046, China; {\it lixd@nju.edu.cn}\\
        \and
             Key laboratory of Modern Astronomy and Astrophysics (Nanjing University), Ministry of Education, Nanjing 210046, China
   }

   \date{Received~~2009 month day; accepted~~2009~~month day}

\abstract{The binding energy parameter $\lambda$ plays a vital role in common envelope evolution. Though it is well known that $\lambda$ takes different values for stars with different masses and varies during stellar evolution, it has been erroneously adopted as a constant in most of the population synthesis calculations. We have systematically calculated the values of $\lambda$ for stars of masses $1-60\,M_{\odot}$ by use of an updated stellar evolution code, taking into account contribution from both gravitational energy and internal energy to the binding energy of the envelope. We adopt the criterion for the core-envelope boundary advocated by \citet{Ivanova2011}. A new kind of $\lambda$ with the enthalpy prescription is also investigated. We present fitting formulae for the calculated values of various kinds of $\lambda$, which can be used in future population synthesis studies.
}

   \authorrunning{Wang, Jia \& Li}            
   \titlerunning{The Binding Energy Parameter $\lambda$ for CE Evolution }  

   \maketitle

\section{Introduction}        
Common envelope (CE) evolution is one of the most important and yet unresolved stages in the formation of various types of binary systems including low-mass X-ray binaries and cataclysmic variables.
For semi-detached binaries, mass transfer can be dynamically unstable if the mass ratio is larger than a critical value or the envelope of the donor star is in convective equilibrium. In this case, the accreting star, usually the less massive star, cannot maintain thermal equilibrium, and the transferred material accumulates on its surface. As a result, both components are expected to overflow their respective Roche lobes (RLs), forming an envelope enshrouding both stars. The accreting star then spirals into the donor's envelope, using its orbital energy to expel the envelope. This is the so-called common envelope (CE) evolution (see \citealt{IbenLivio1993,TaamSandquist2000,Ivanova2013} for reviews). The outcome of CE evolution is either a compact binary consisting of the donor's core and the companion star, or a single object due to merger of the two stars, depending on whether the available orbital energy is large enough to eject the donor's envelope. This process can be described by the following equation (\citealt{Webbink1984}),
 \begin{equation}
   E_{\rm bind}=\alpha_{\rm CE}\left(\frac{GM_{\rm core}M_2 }{2a_{\rm f} }
   -\frac{GM_1 M_2}{2a_{\rm i}}\right),
 \end{equation}
where
\begin{equation}
E_{\rm bind}= -\int_{M_{\rm core}}^{M_1}\frac{GM(r)}{r}dm
 \end{equation}
is the binding energy of the envelope, $\alpha_{\rm CE}$ the efficiency parameter that denotes the fraction of the orbital energy used to eject the CE, $G$ the gravitational constant, $M_{1}$ and $M_{\rm core}$ the masses of the donor and its core, $M_{2}$ the mass of the companion star, and $a_{\rm i}$ and $a_{\rm f}$ the pre- and post-CE orbital separations, respectively.

It has been suggested that the internal energy (including both thermal and recombination energies) in the envelope may also contribute to the binding energy, so a more general form for $E_{\rm bind}$ can be written as
\begin{equation}
E_{\rm bind}= \int_{M_{\rm core}}^{M_1}\left[-\frac{GM(r)}{r}+U\right ] dm,
 \end{equation}
where $U$ is the internal energy (\citealt{Han1994,DewiTauris2000}). More recently,
\cite{IvanovaChaichenets2011} proposed that this canonical energy formalism should be modified with an additional $P/\rho$ term (where $P$ is the pressure and  $\rho$ is the density of the gas) by
taking into account the mass outflows during the spiral-in stage. These authors argue that, the standard form (1) or (2) is based on the consideration that the envelope of a giant star is dispersed or unstable once its total energy $W_{\rm env}>0$, but neither of the two considerations has to occur when the envelope has quasi-steady outflows. For such envelopes, the material obeys the first law of thermodynamics, and the criterion for a mass shell to reach the point of no return in its expansion turns to be that the sum of its kinetic energy, potential energy, and enthalpy,  rather than the total energy, becomes positive.  This is so-called the enthalpy model  (see \citealt{Ivanova2013} for a detailed discussion). Assuming that the velocity of gas at infinity is zero, the binding energy is expressed as
\begin{equation}
    E_{\rm bind}= \int_{M_{\rm core}}^{M_1}\left[-\frac{GM(r)}{r}+U+\frac{P}{\rho}\right] dm.
\end{equation}
Since the $P/\rho$ term is always non-negative and orders of magnitude larger than $U$, the
absolute value of $E_{\rm bind}$ decreases substantially in this case. One should be cautious that  quasi-stationary mass outflow only develops when the envelope experiences a slow self-regualted phase during the spiral-in stage, that is, the spiral-in phase proceeds on a thermal timescale (\citealt{Ivanova2013}).

\cite{deKool1990} proposed a convenient way to evaluate the binding energy by introducing a parameter $\lambda$ to characterizing the central concentration of the donor's envelope,
\begin{equation}
    E_{\rm bind} = -\frac {GM_1 M_{\rm env}}{\lambda a_{\rm i} r_{\rm L}},
\end{equation}
where $M_{\rm env}=M_1-M_{\rm core}$ is the mass of the envelope,
$r_{\rm L}= R_{\rm L}/ a_{\rm i}$ is the ratio of the donor's RL
radius and the orbital separation at the onset of CE. Typically,
$a_{\rm i}r_{\rm L}$ is taken to
be the stellar radius once a star fills its RL. Thus the post-CE separation can be determined by inserting Eq.~(5) into Eq.~(1),
\begin{equation}
    \frac{a_{\rm f}}{a_{\rm i}}= \frac{M_{\rm core} M_{\rm 2}}{M_1}
    \frac{1}{M_{\rm 2}+2M_{\rm env}/\alpha_{\rm CE} \lambda r_{\rm L}}.
\end{equation}

It should be emphasized that both $\alpha_{\rm CE}$ and $\lambda$ are variables depending on stellar and binary parameters, although they have been treated as constant ($<1$) in most of the population synthesis calculations, due to both poor understanding of them and convenience for calculation. However, many studies (e.g. \citealt{DewiTauris2000, Podsiadlowski2003, Webbink2008, XuLi2010a, XuLi2010b,Wong2014})
have shown that $\lambda$ varies as the star evolves and can deviate far from a constant value (say, 0.5). Some investigations also suggested that $\alpha_{\rm CE}$ may depend on the binary parameters such as the component mass and the orbital period (e.g., \citealt{TaamSandquist2000, Podsiadlowski2003, DeMarco2011, Davis2012}).

Systematic calculations of the values of $\lambda$ have been performed by \citet{DewiTauris2000},
\citet{Podsiadlowski2003}, and \citet{XuLi2010a, XuLi2010b}. In the latter, fitting formulae for $\lambda$ have also been provided so they can be incorporated into population synthesis investigations. In this work we re-visit this problem and provide more reliable $\lambda$ values by taking into account the following factors.

First, we adopt the Modules for Experiments in Stellar Astrophysics (MESA) code (\citealt{Paxton2011, Paxton2013, Paxton2015}) to calculate stellar evolution, which is more powerful in probing the stellar structure than \citet{Eggleton1971}'s evolution code EV previously adopted by \citet{XuLi2010a, XuLi2010b}. Employing modern software engineering tools and techniques
allows MESA to consistently evolve stellar models through challenging phases for stellar evolution codes in the past, for example, the He core flash in low-mass stars and advanced nuclear
burning in massive stars (\citealt{Paxton2011}). It also adopts denser grids for stellar structure than the EV code. We find that stars appear to be generally less compact after evolving off main sequence when modeled with MESA compared with the EV code. The structure of the hydrogen-burning shell, which is near the defined core-envelope boundary, plays a vital role in determining the value of $\lambda$.

Second, besides the traditional $\lambda$ related to gravitational energy and internal energy, we also calculate the values of $\lambda$ in the enthalpy prescription.

Third, it is well known that the $\lambda$-value is sensitive to the definition of the core-envelope boundary (see \citealt{Ivanova2013} for a detailed discussion). It was arbitrarily assumed to be the $(10-15)\%$ hydrogen layer in \citet{DewiTauris2000} and \citet{XuLi2010a}. \cite{Ivanova2011} proposed that this boundary should be defined in the hydrogen shell which has the maximum local sonic velocity (i.e., the maximal compression) prior to CE evolution. This criterion comes from the study of the outcome of the CE event and the fact that a He core would experience a post-CE thermal readjustment phase, and presents a more self-consistent definition of the core-envelope boundary.

This paper is organized as follows. In section 2, we briefly describe the stellar models and assumptions adopted. We present the calculated results and fitting formulae for $\lambda$ in section 3. Our conclusions are in section 4.

\begin{figure}
     \begin{tabular}{cccc}
     \begin{minipage}[h,t]{1.9in}
     \includegraphics[width=1.9in]{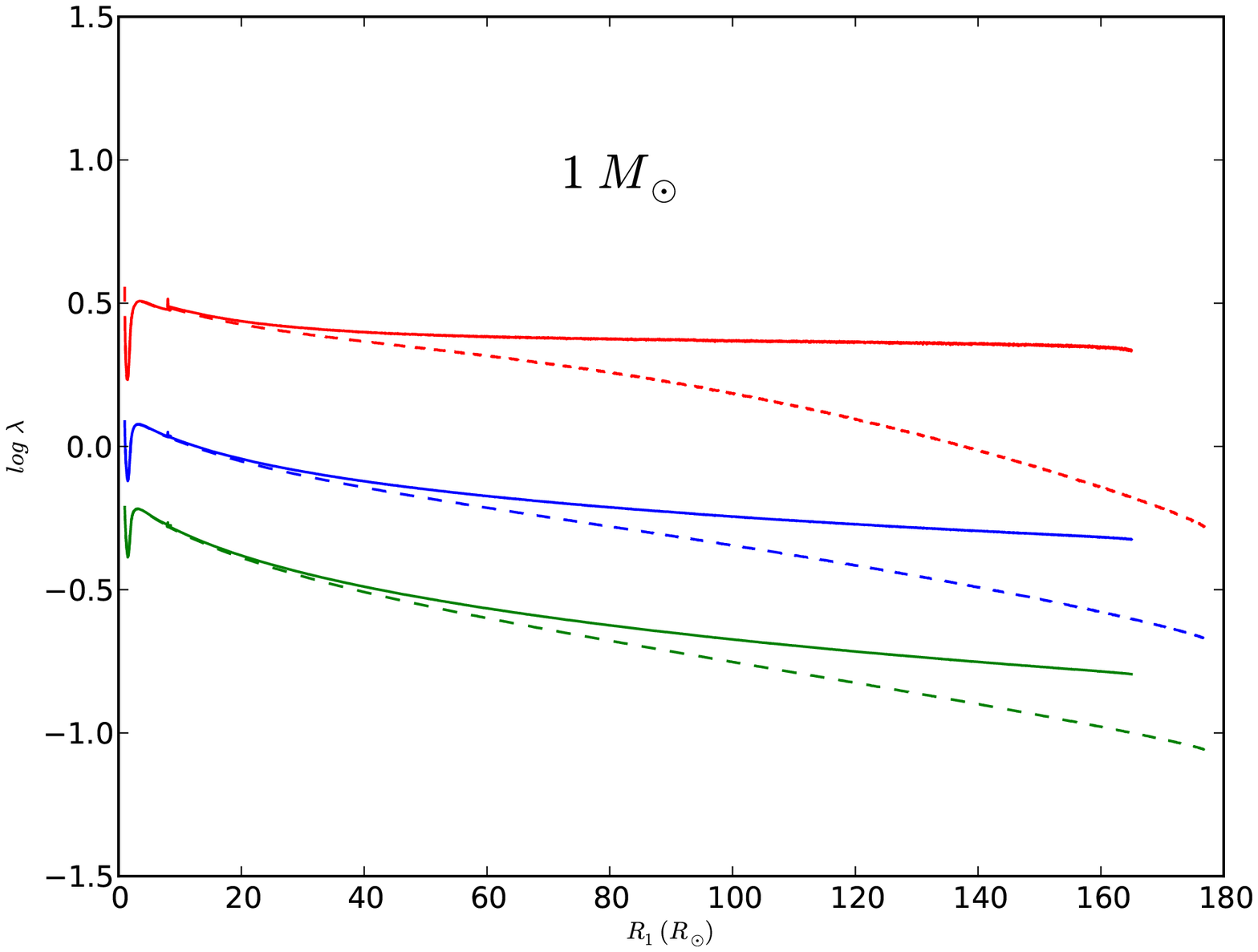}
     \end{minipage}
     \begin{minipage}[h,t]{1.9in}
     \includegraphics[width=1.9in]{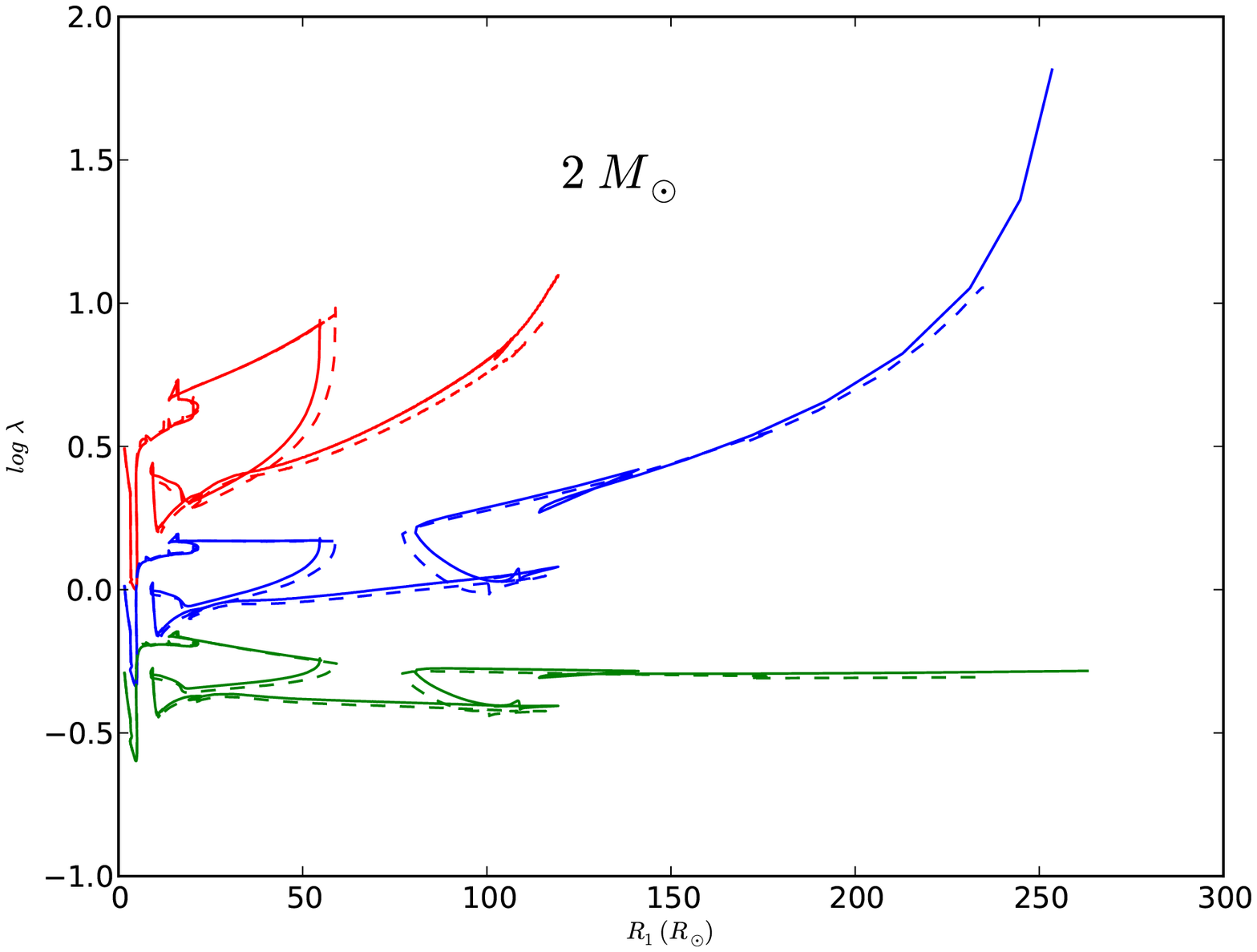}
     \end{minipage}
     \begin{minipage}[h,t]{1.9in}
     \includegraphics[width=1.9in]{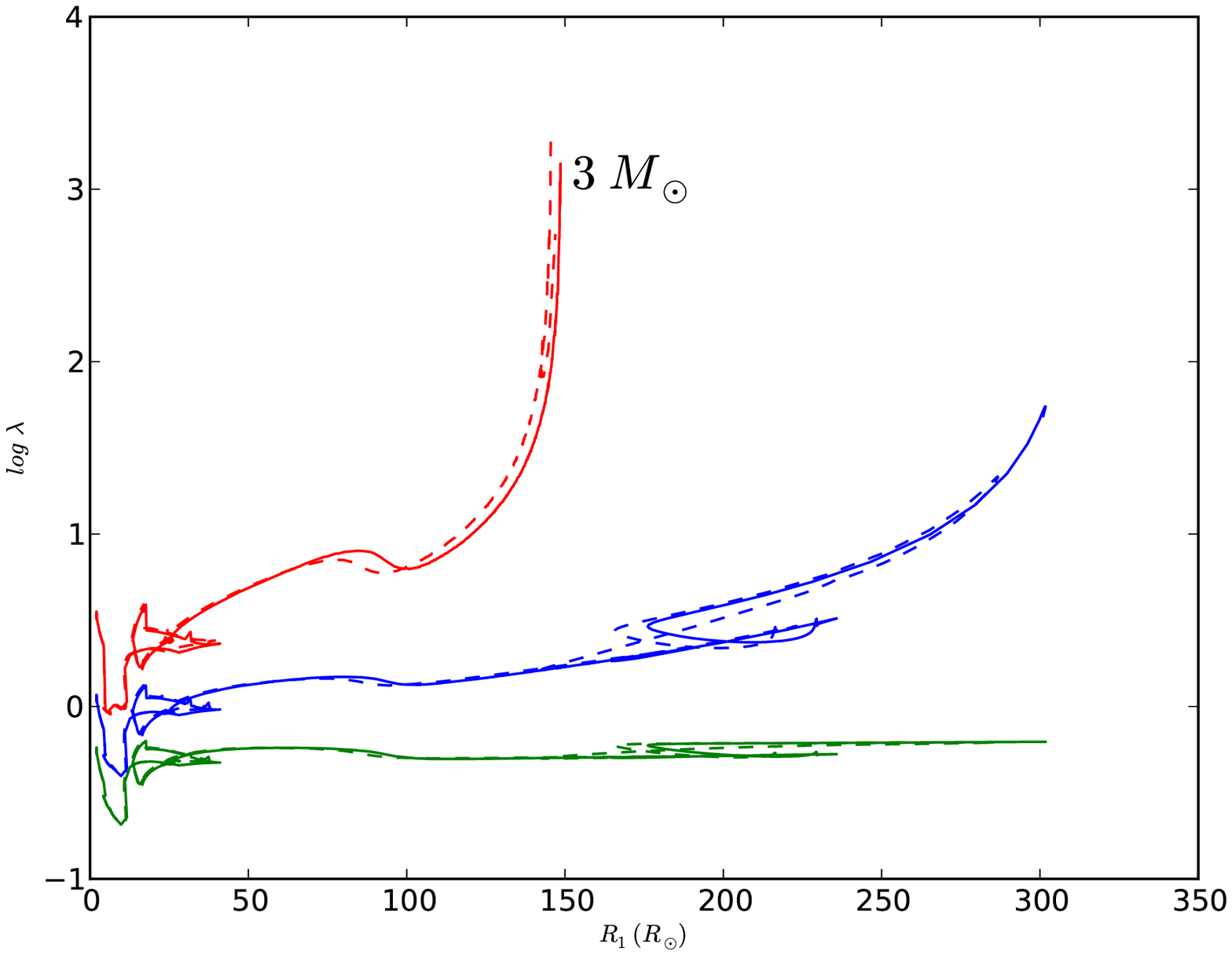}
     \end{minipage}\\
     \begin{minipage}[h,t]{1.9in}
     \includegraphics[width=1.9in]{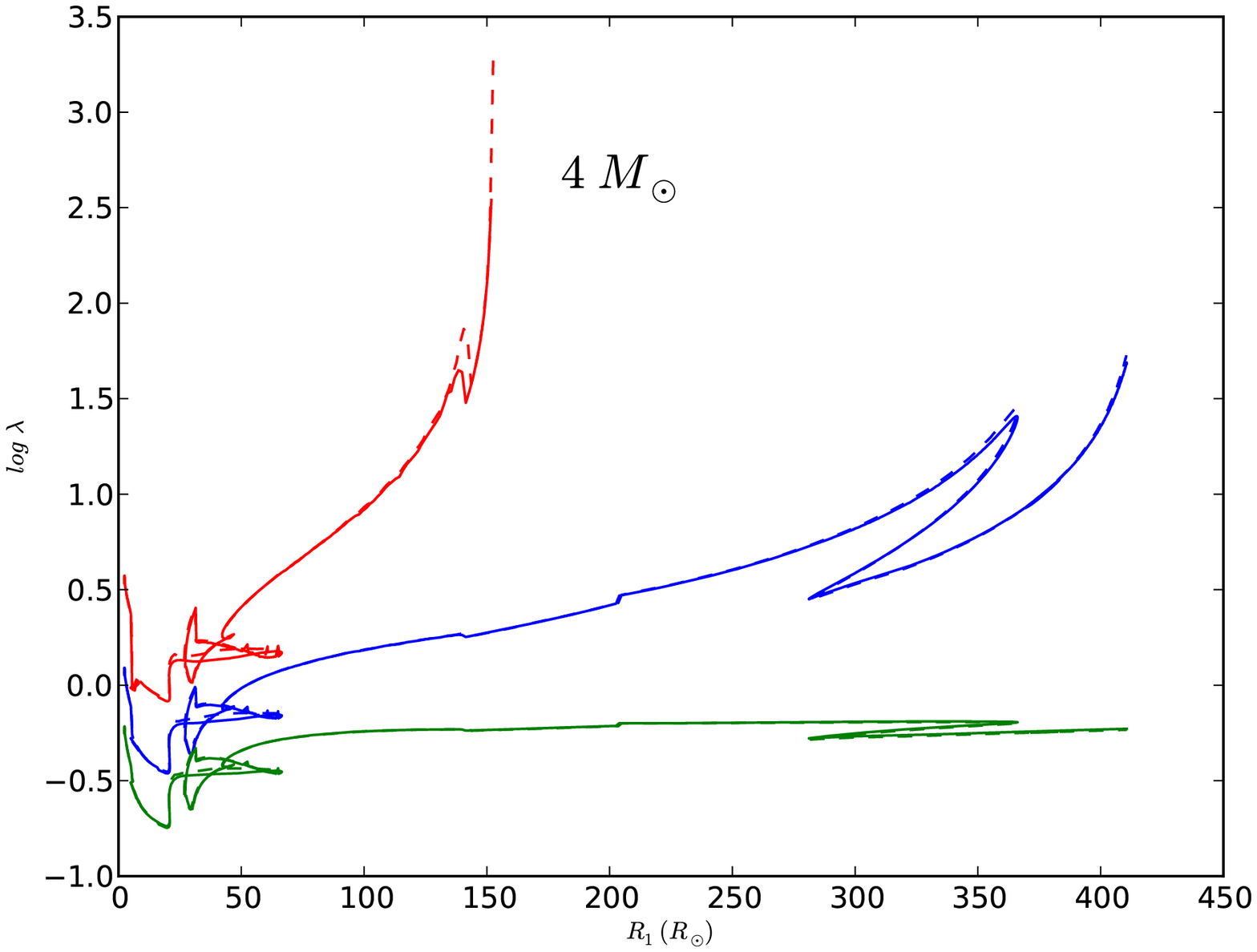}
     \end{minipage}
     \begin{minipage}[h,t]{1.9in}
     \includegraphics[width=1.9in]{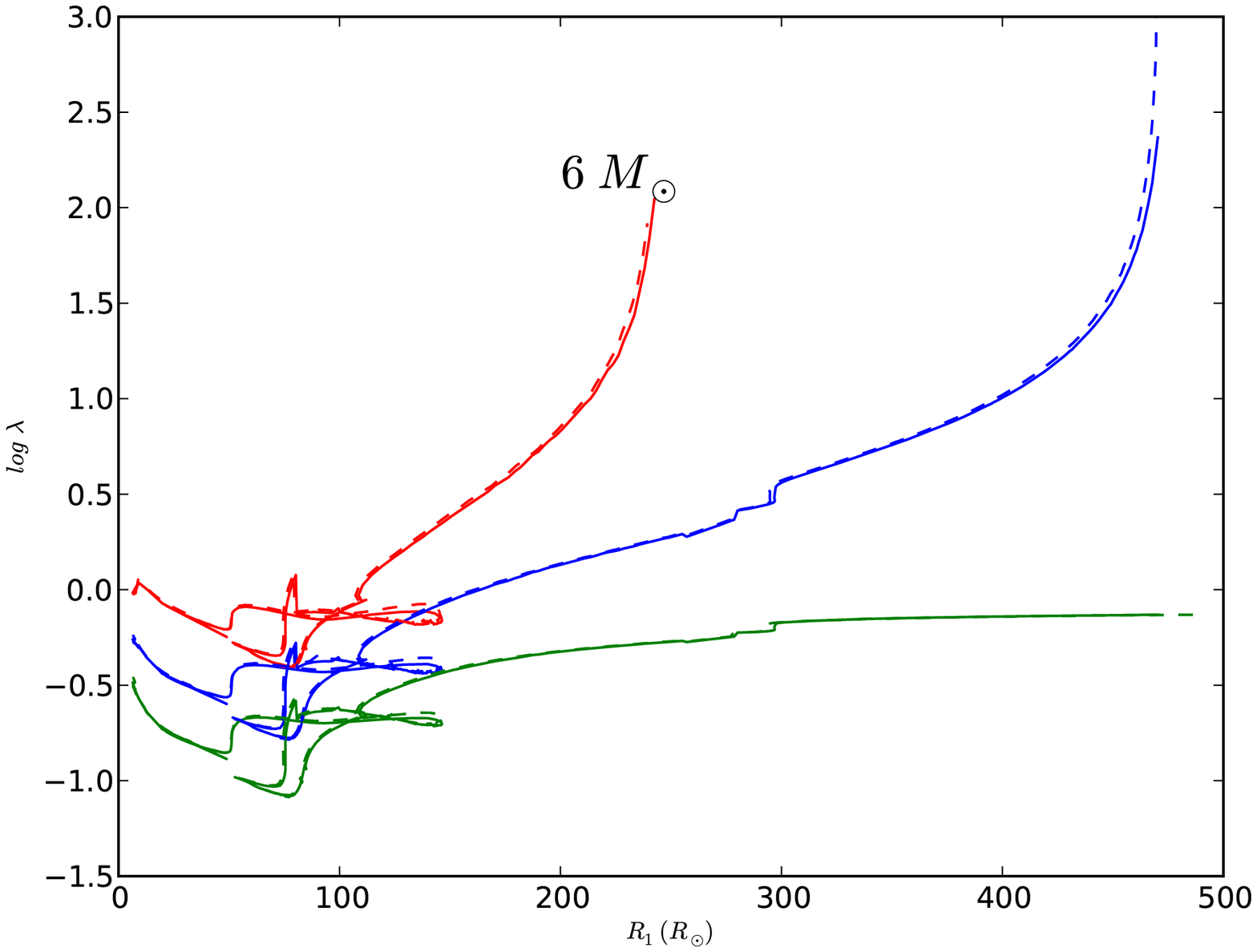}
     \end{minipage}
     \begin{minipage}[h,t]{1.9in}
     \includegraphics[width=1.9in]{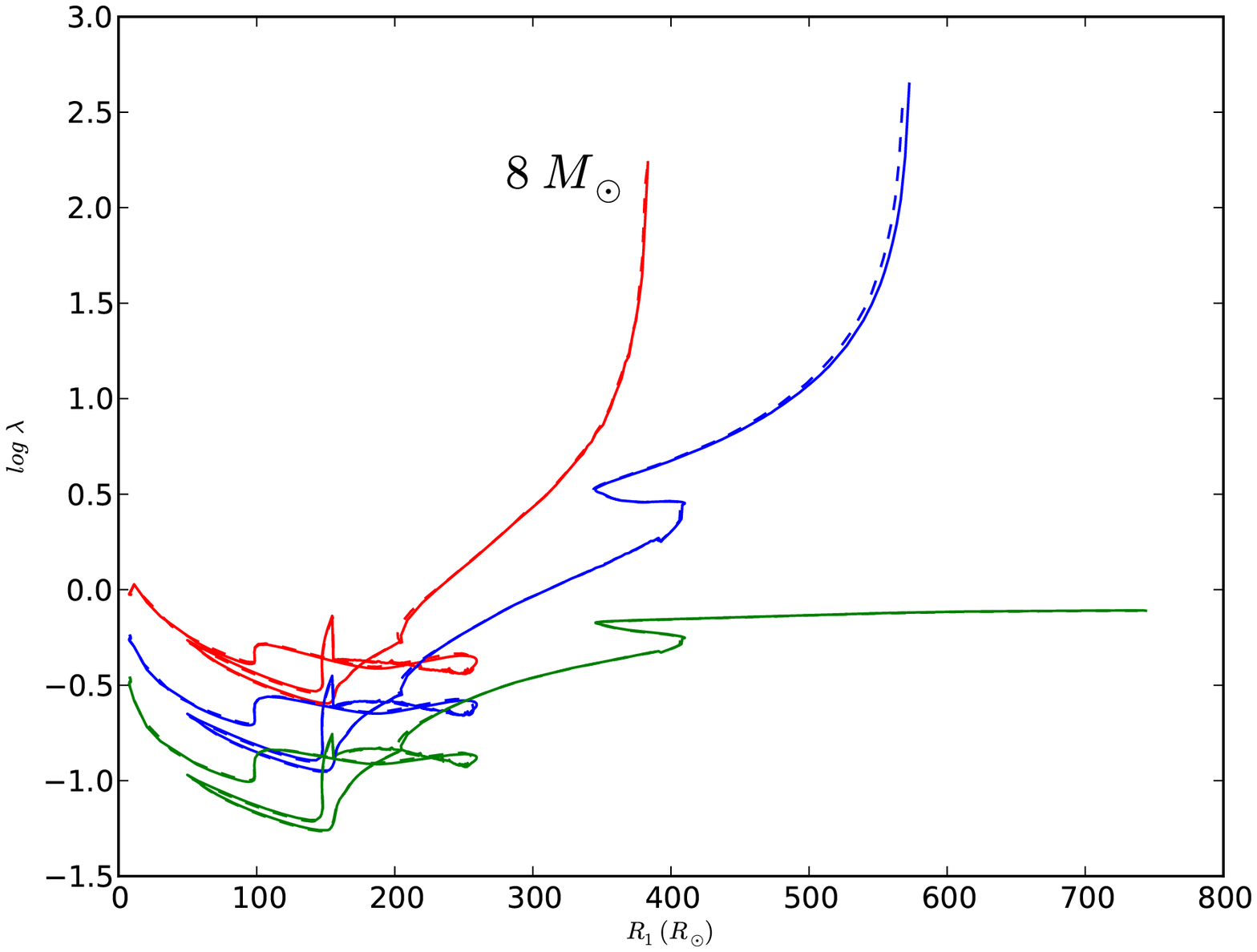}
     \end{minipage}\\
     \begin{minipage}[h,t]{1.9in}
     \includegraphics[width=1.9in]{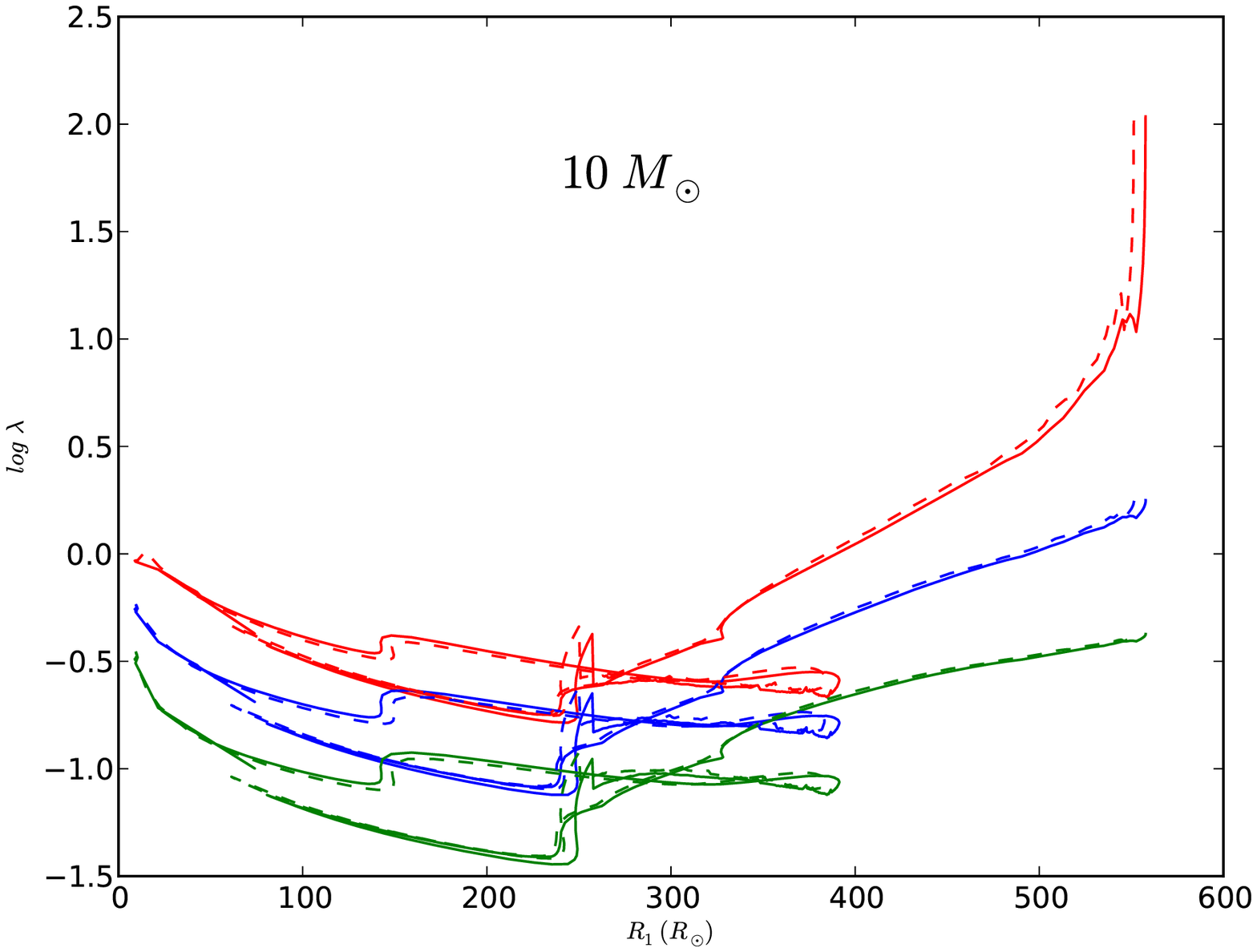}
     \end{minipage}
     \begin{minipage}[h,t]{1.9in}
     \includegraphics[width=1.9in]{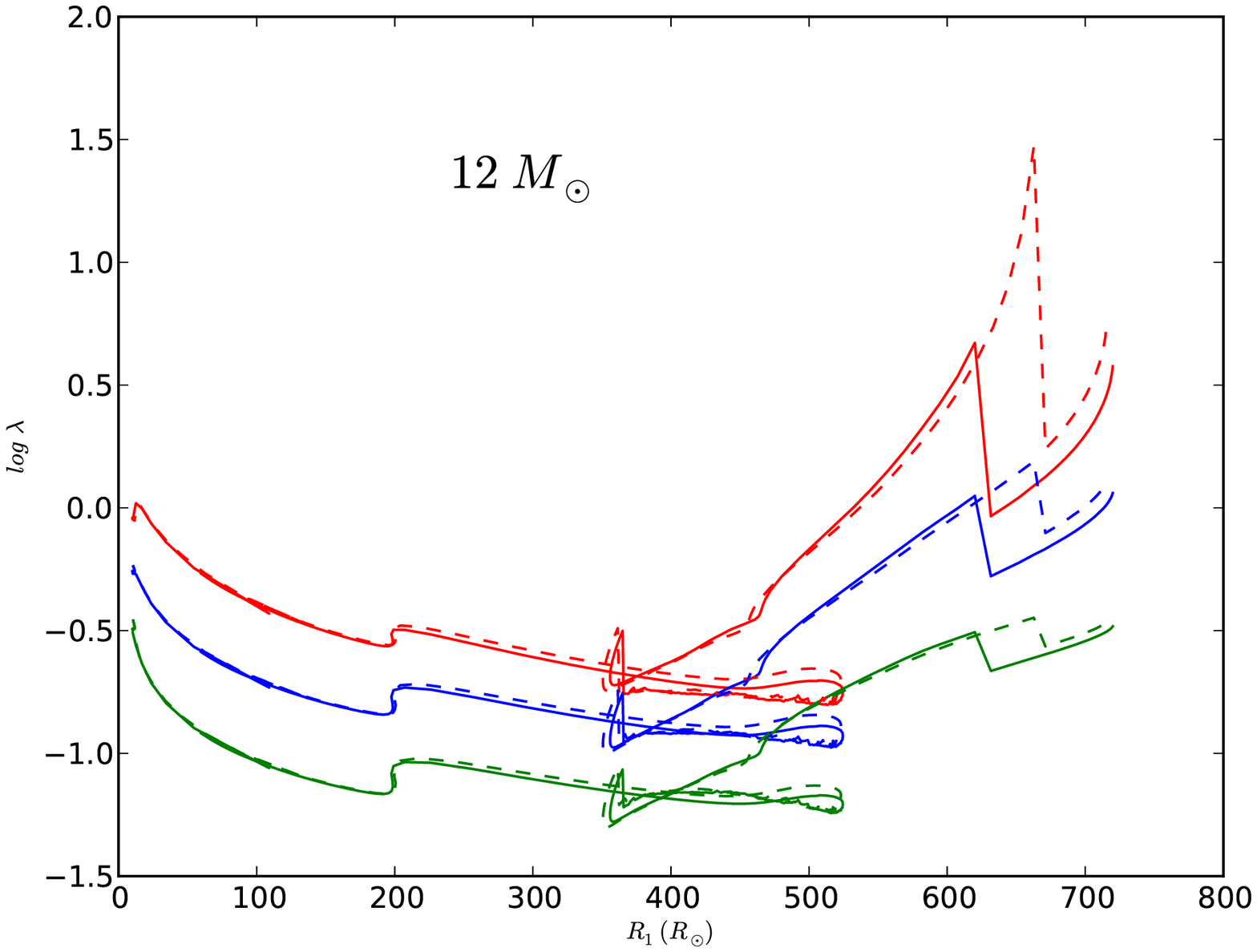}
     \end{minipage}
     \begin{minipage}[h,t]{1.9in}
     \includegraphics[width=1.9in]{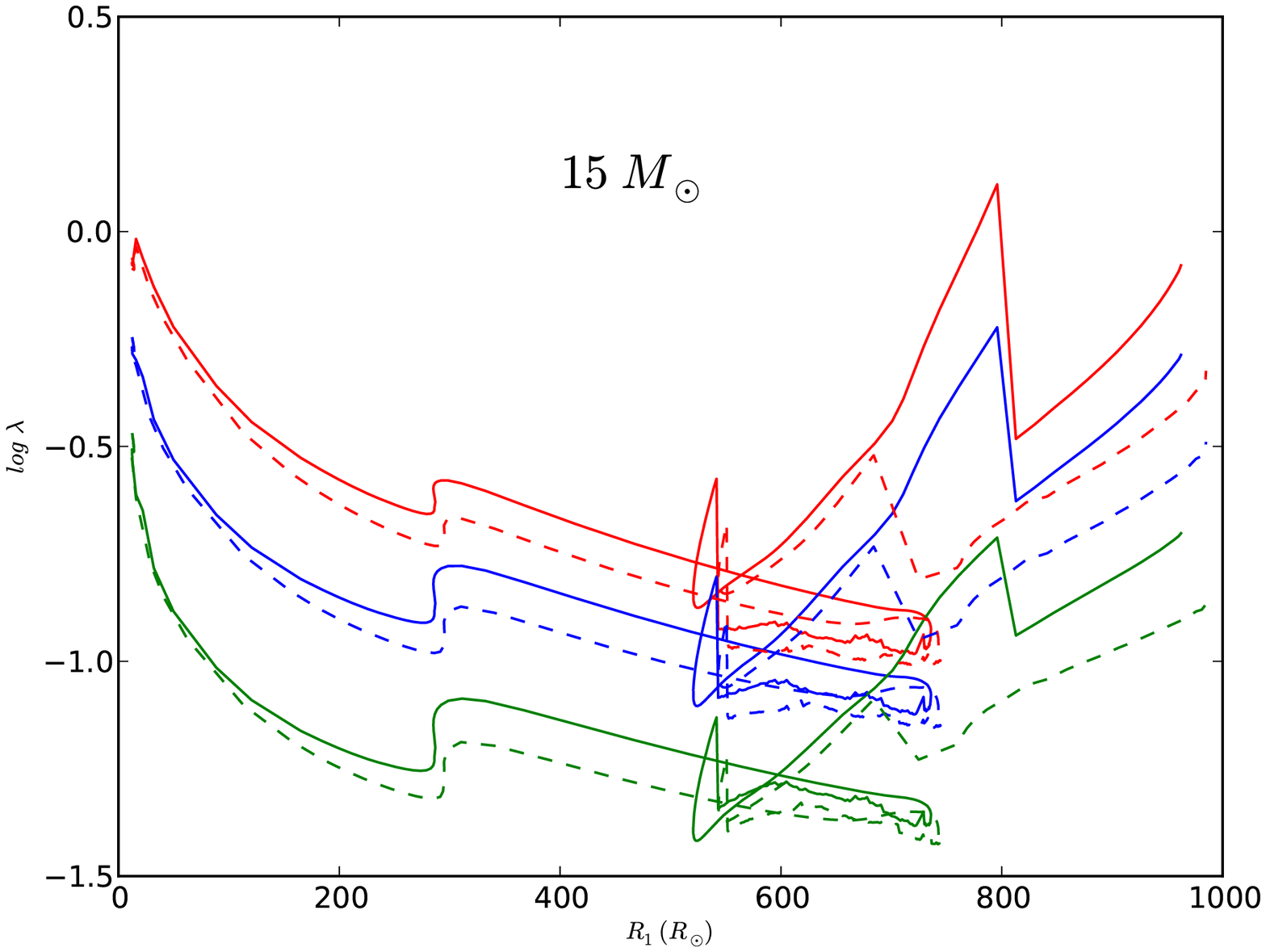}
     \end{minipage}
     \end{tabular}
\caption{Evolution of the binding energy parameters $\lambda$
with the stellar radius $R$ for $1-15\,M_\odot$ stars. The red, blue and green lines represent $\lambda_{\rm h}$,
$\lambda_{\rm b}$, and $\lambda_{\rm g}$, and the solid and dashed lines
represent the results with the Wind1 and Wind2 prescriptions, respectively. \label{figure1}}
\end{figure}

\begin{figure}
    \begin{tabular}{ccc}
     \begin{minipage}[h,t]{1.9in}
     \includegraphics[width=1.9in]{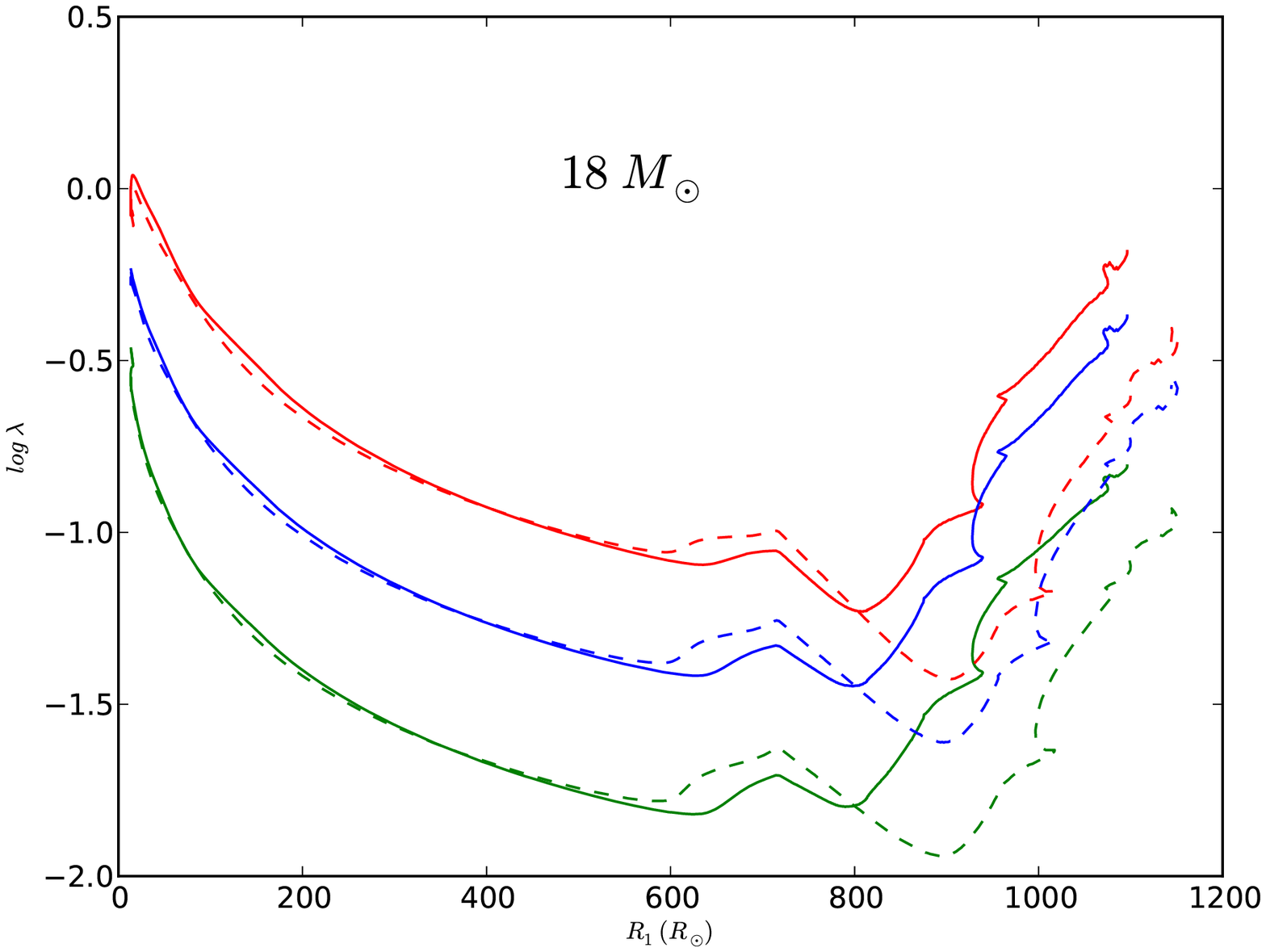}
     \end{minipage}
     \begin{minipage}[h,t]{1.9in}
     \includegraphics[width=1.9in]{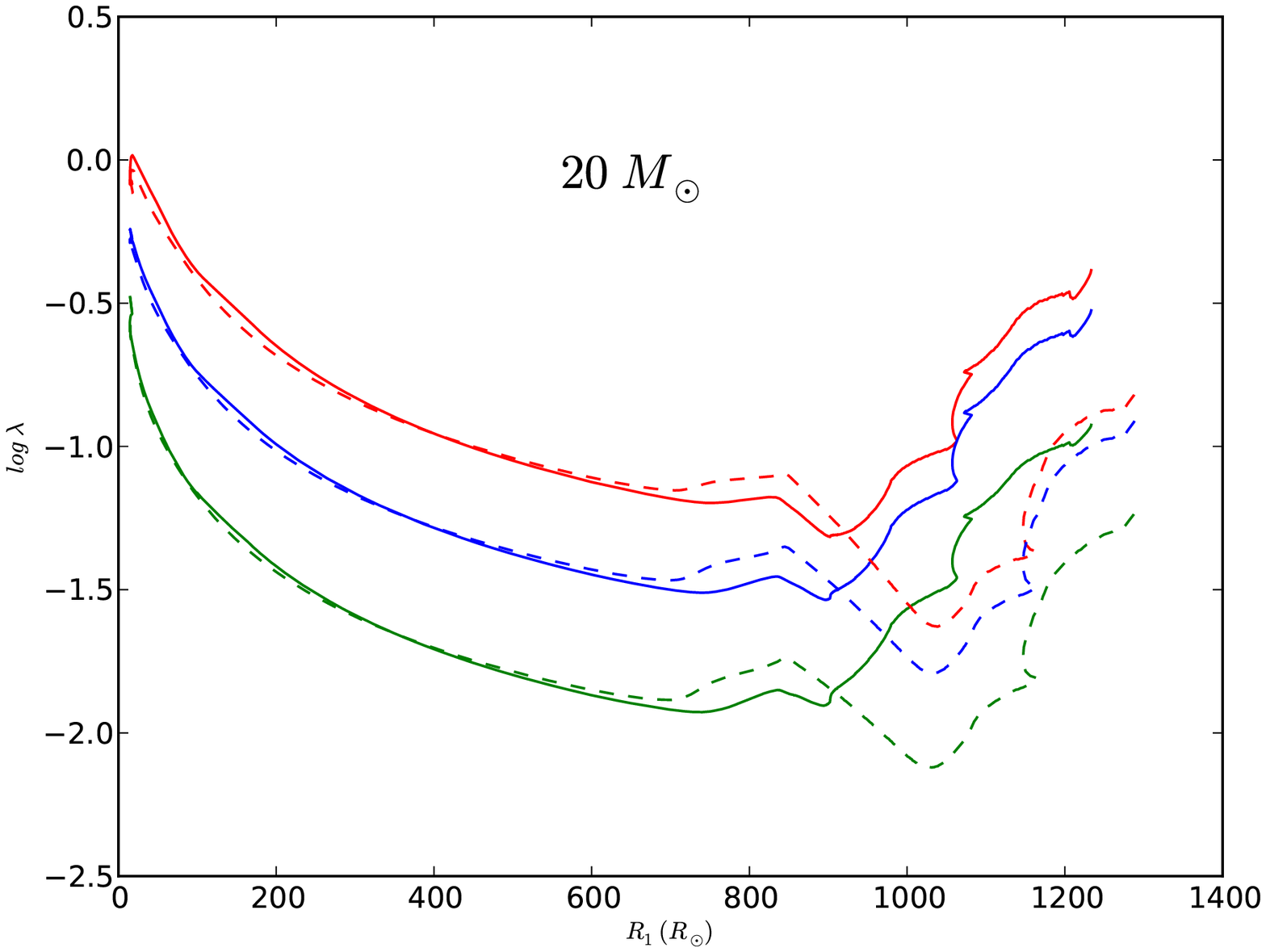}
     \end{minipage}
     \begin{minipage}[h,t]{1.9in}
     \includegraphics[width=1.9in]{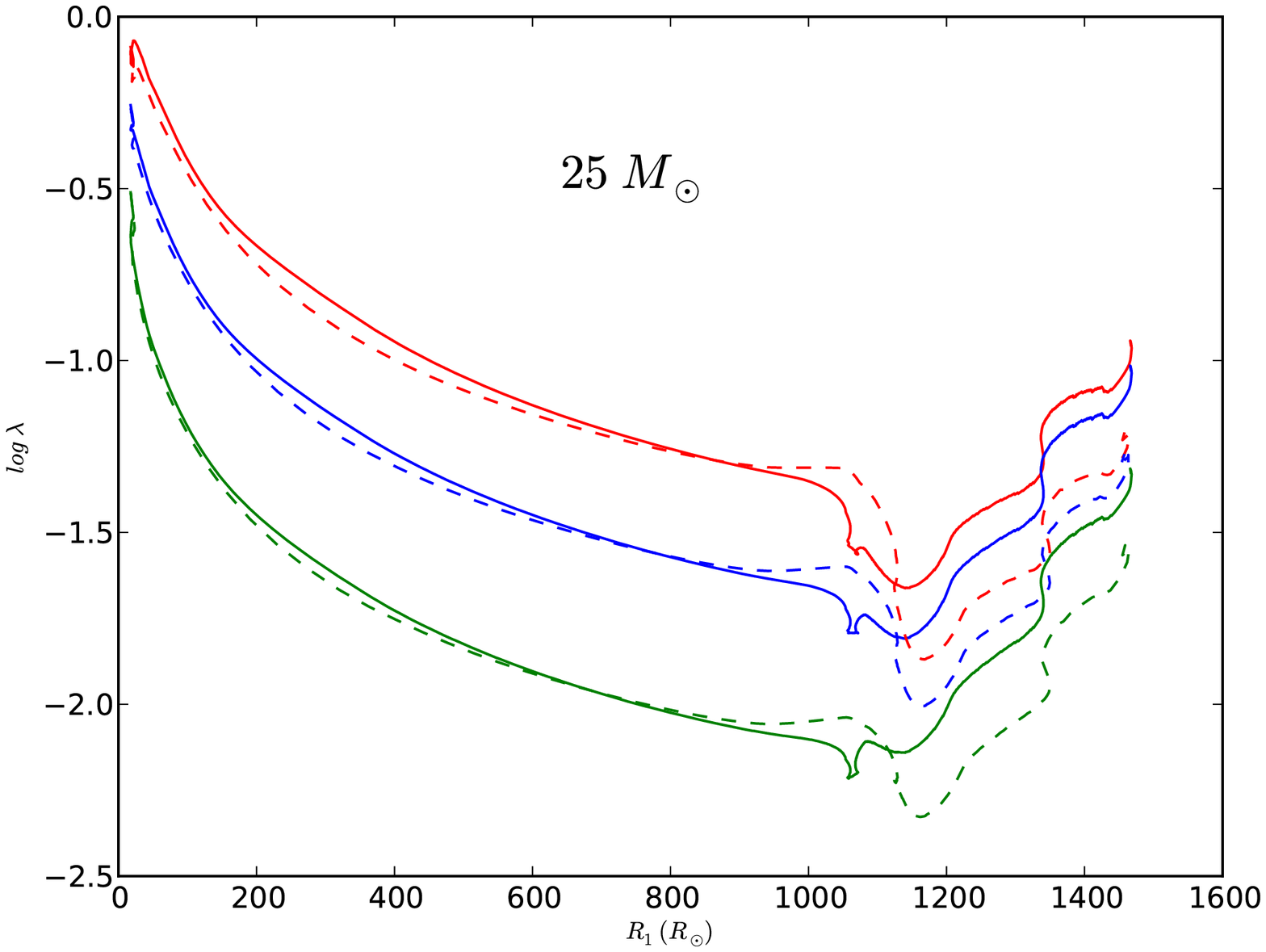}
     \end{minipage} \\
     \begin{minipage}[h,t]{1.9in}
     \includegraphics[width=1.9in]{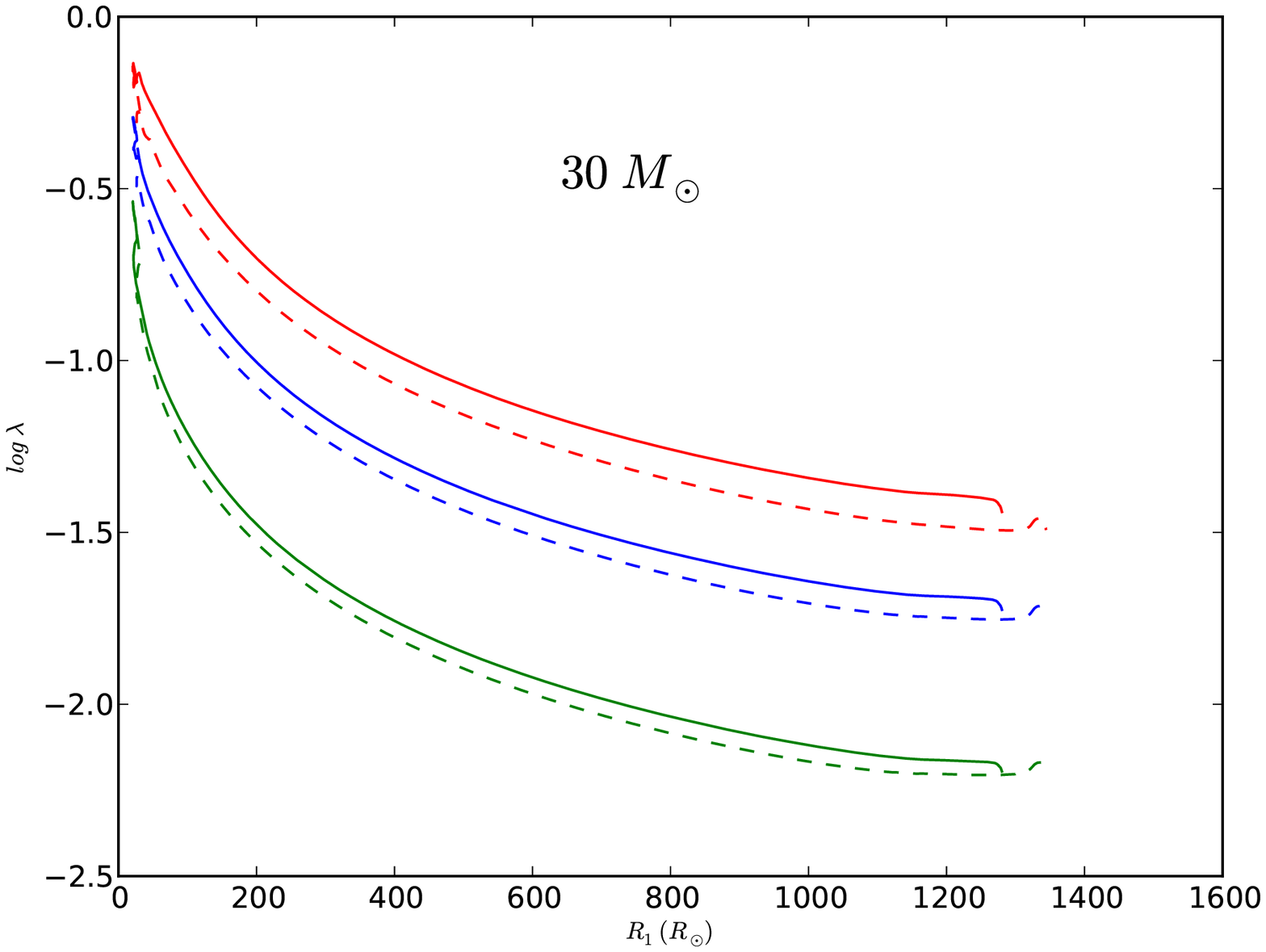}
     \end{minipage}
     \begin{minipage}[h,t]{1.9in}
     \includegraphics[width=1.9in]{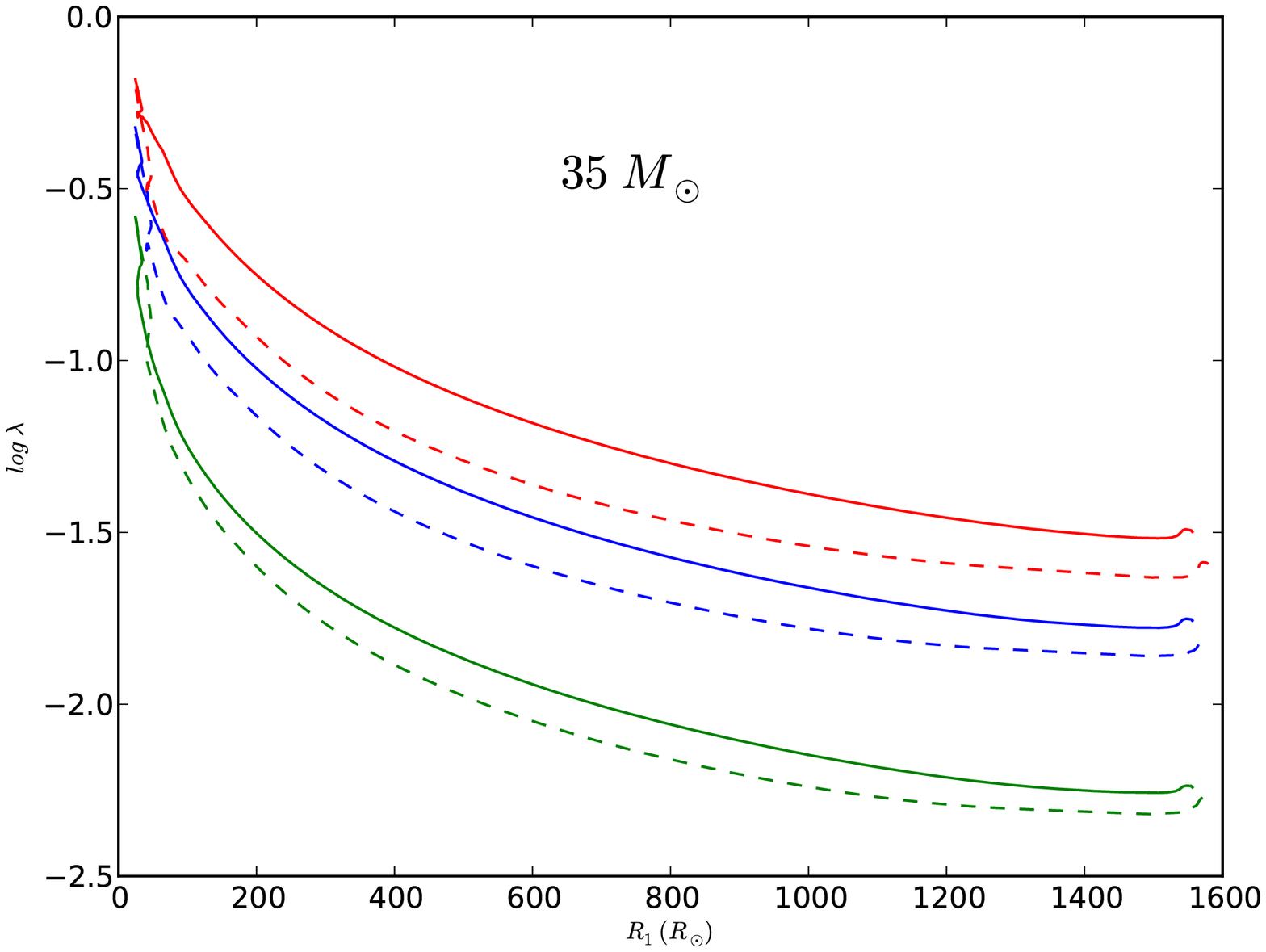}
     \end{minipage}
     \begin{minipage}[h,t]{1.9in}
     \includegraphics[width=1.9in]{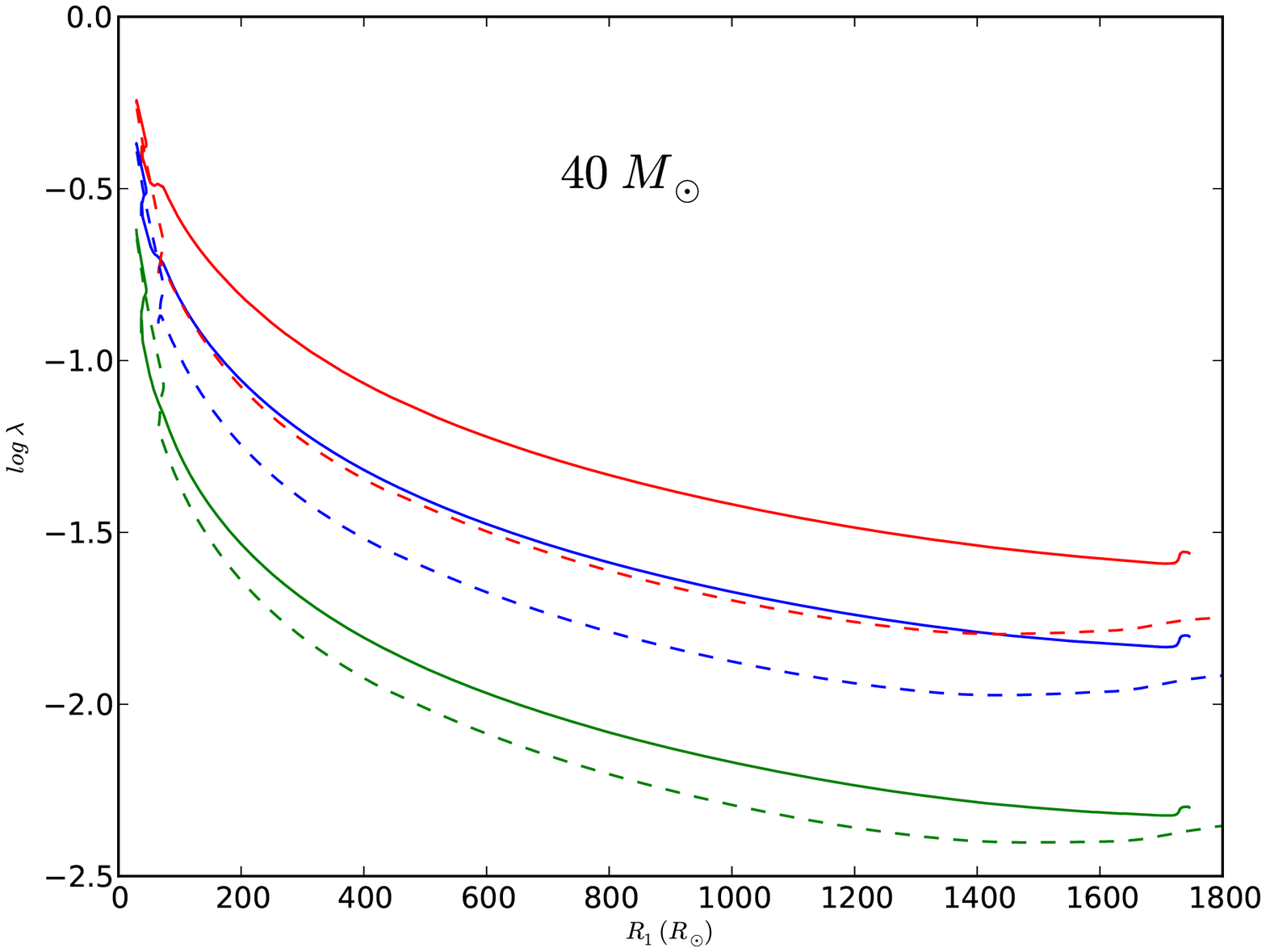}
     \end{minipage}\\
     \begin{minipage}[h,t]{1.9in}
     \includegraphics[width=1.9in]{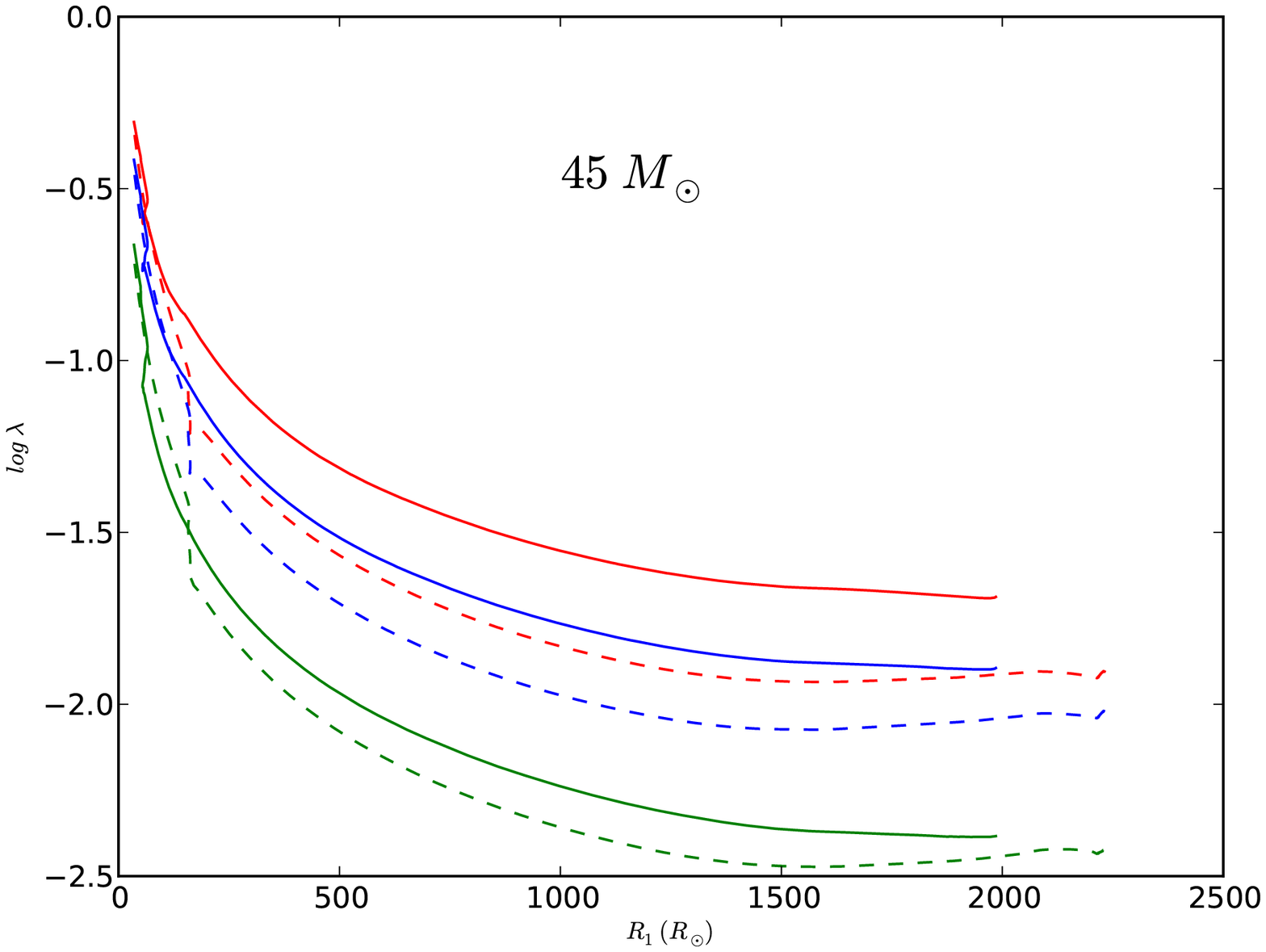}
     \end{minipage}
     \begin{minipage}[h,t]{1.9in}
     \includegraphics[width=1.9in]{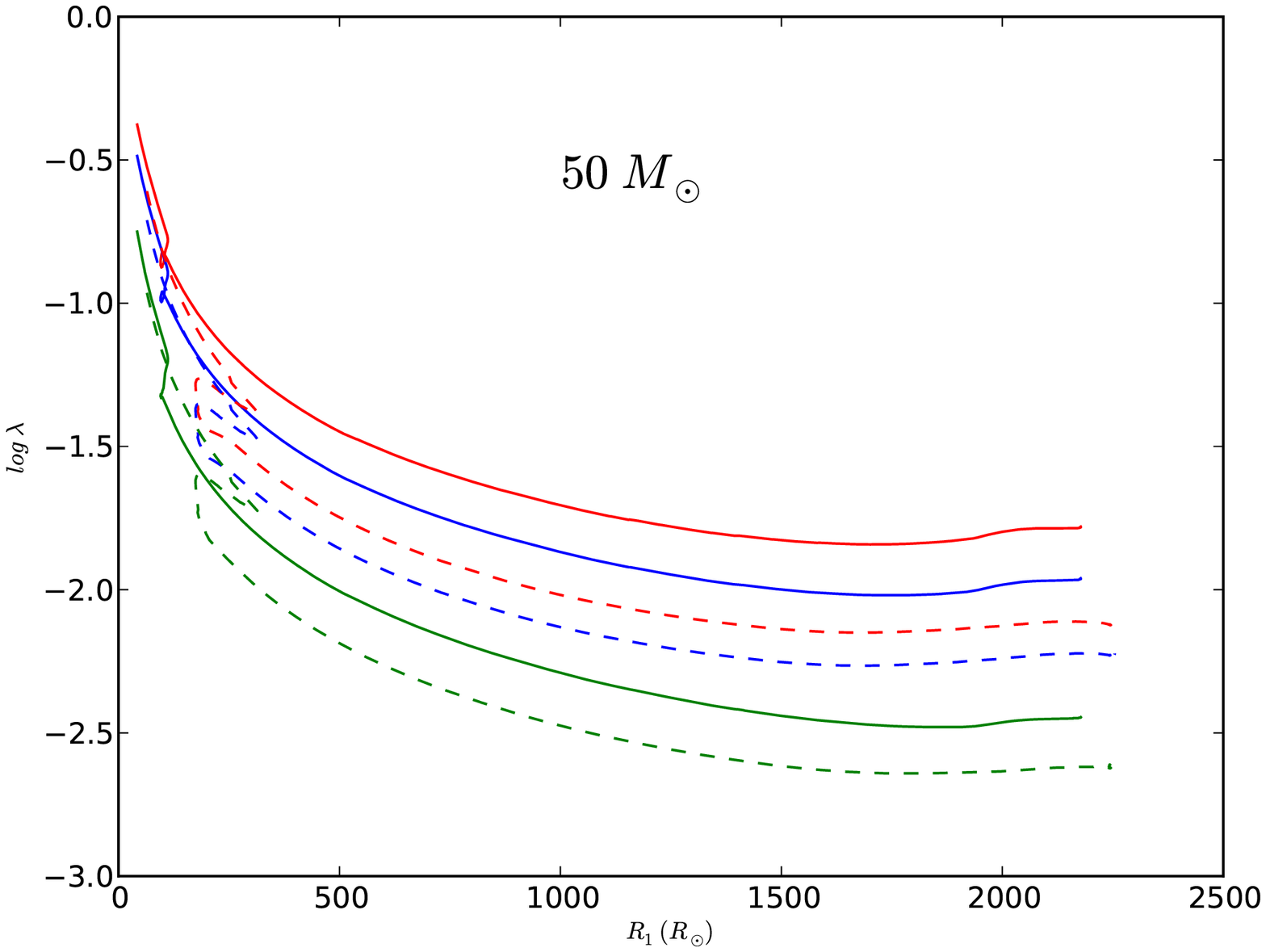}
          \end{minipage}
     \begin{minipage}[h,t]{1.9in}
     \includegraphics[width=1.9in]{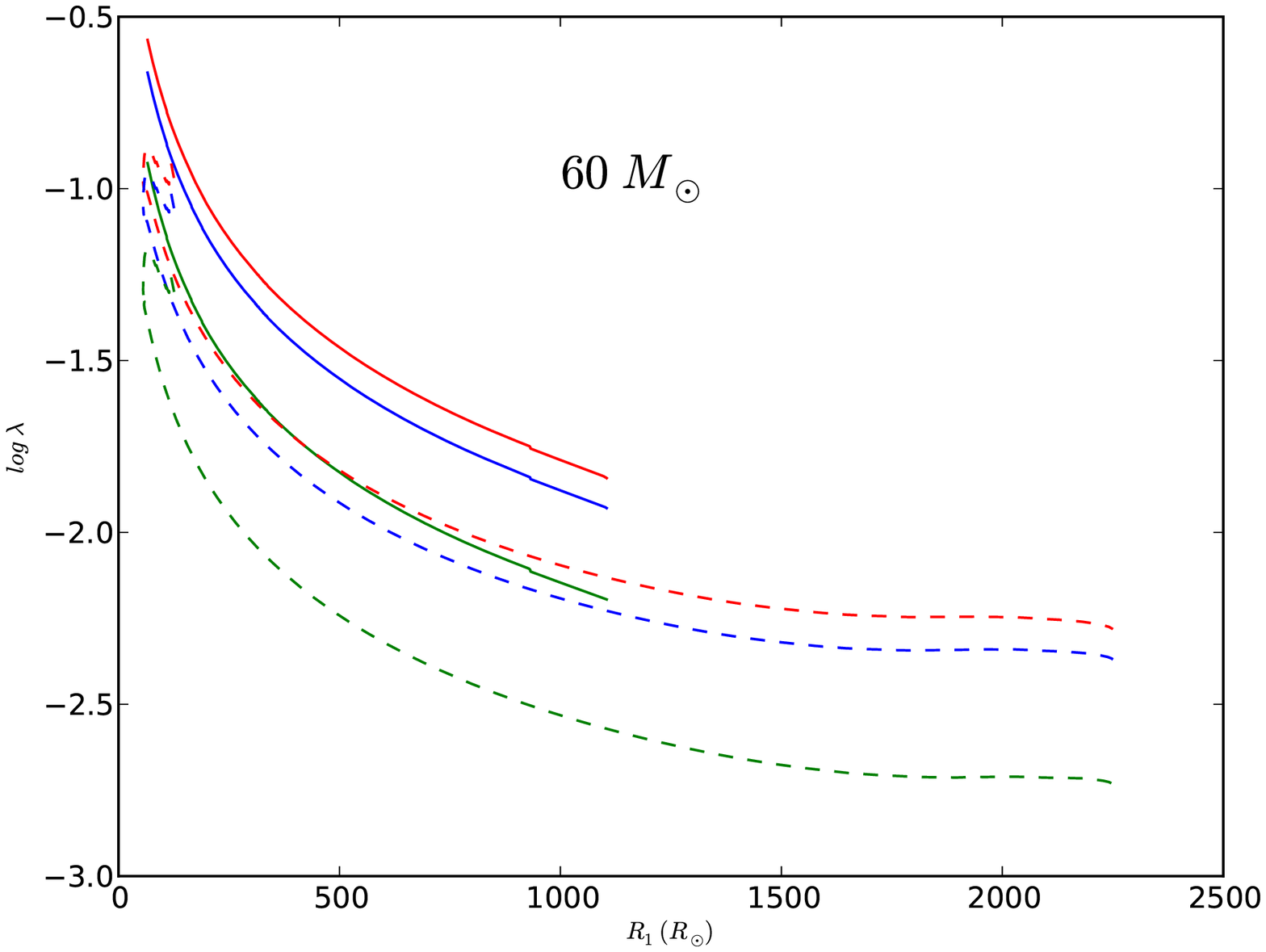}
     \end{minipage}
     \end{tabular}
\caption{Same as Fig.~\ref{figure1} but for $18-60\,M_\odot$ stars. \label{figure2}}
\end{figure}

\section{Model Description}
We adopted an updated version (7624) of the MESA code to calculate the binding energy parameter $\lambda$ for stars with initial masses in the range of $1-60\, M_{\sun}$. We consider Pop. I stars with the chemical compositions of $X=0.7$ and $Z=0.02$. Our previous study has shown that there is not significant change in the values of $\lambda$ for Pop. I and II stars (\citealt{XuLi2010a, XuLi2010b}).

It has been recognized that stellar winds play an important role in determining the $\lambda$ parameter, especially for massive stars (\citealt{Podsiadlowski2003}). Here, we adopt two prescriptions for the wind mass loss rates. The first one, denoted as Wind1, is same as in  \cite{Hurley2000} and \cite{Vink2001} (for O and B stars), and the second, denoted as Wind2, takes the maximum value of the above loss rates in all the evolutionary stages, to be consistent with \cite{XuLi2010a, XuLi2010b}.

The Wind1 prescription is described as follows:\\
(1) Stellar wind mass loss described by \cite{NieuwenhuijzendeJager1990}:
\begin{equation}
\dot{M}_{\rm NJ}(M_{\odot}{\rm yr}^{-1})=9.6\times 10^{-15}R^{0.81}L^{1.24}M^{0.16},
\end{equation}
where $M$, $R$, and $L$ are the stellar mass, radius, and luminosity in solar units, respectively.\\
(2) Wind loss from giant branch stars by \cite{KudritzkiReimers1978}:
\begin{equation}
\dot{M}_{\rm R}(M_{\odot}{\rm yr}^{-1})=2\times 10^{-13}\frac{LR}{M}.
\end{equation}
(3) Wind loss from AGB stars by \cite{VassiliadisWood1993}:
\begin{equation}
{\rm log}\, \dot{M}_{\rm VW}(M_{\odot}{\rm yr}^{-1})=-11.4+0.0125[P_0-100\, {\rm max}(M-2.5,0.0)],
\end{equation}
where
\begin{equation}
{\rm log}\, P_0={\rm min}(3.3,-2.07-0.9\, {\rm log}\, M+ 1.94\, {\rm log}\, R).
\end{equation}
The maximum wind loss rate in this prescription is limited to
\begin{equation}
\dot{M}_{\rm VW,max}=1.36\times 10^{-9}L\,M_{\odot}{\rm yr}^{-1}.
\end{equation}\\
(4) Wolf-Rayet-like mass loss by \cite{Hamann1995} and \cite{HamannKoesterke1998}:
\begin{equation}
\dot{M}_{\rm WR}(M_{\odot}{\rm yr}^{-1})=10^{-13}L^{1.5}(1.0-\mu)M_{\odot}{\rm yr}^{-1},
\end{equation}
with
\begin{equation}
\mu=(\frac{M-M_{\rm core}}{M}){\rm min}\{5.0,{\rm max}[1.2,(\frac{L}{L_0})^\kappa]\},
\end{equation}
where $L_0=7.0\times 10^4$, and $\kappa=-0.5$.\\
(5) O and B type star's wind loss according to \cite{Vink2001}:
\begin{equation}
\begin{split}
{\rm log}\, \dot{M}_{\rm OB}(M_{\odot}{\rm yr}^{-1})= & -6.697(\pm 0.061)+2.194(\pm 0.021)\, {\rm log}\,  (L/10^5)-1.313(\pm 0.046)\, {\rm log} \, (M/30) \nonumber\\
& -1.226(\pm 0.037)\, {\rm log}\,  (\frac{v_{\infty}/v_{\rm esc}}{2.0})+0.933(\pm 0.064)\, {\rm log}\,  (T_{\rm eff}/40000) \nonumber\\
& -10.92(\pm 0.90)\, {\rm log}\,  (T_{\rm eff}/40000)\}^2,
\end{split}
\end{equation}
with $v_{\infty}/v_{\rm esc}=2.6$ for 27500 K $<T_{\rm eff}\leq 50000\,\rm K$ (where $T_{\rm eff}$ is the effective temperature).
\begin{equation}
\begin{split}
{\rm log}\, \dot{M}_{\rm OB}(M_{\odot}{\rm yr}^{-1})= & -6.688(\pm 0.080)+2.210(\pm 0.031)\,
{\rm log}\, (L/10^5)-1.339(\pm 0.068)\, {\rm log}\, (M/30) \nonumber\\
& -1.601(\pm 0.055)\, {\rm log}\,  (\frac{v_{\infty}/v_{\rm esc}}{2.0})+1.07(\pm 0.10)\, {\rm log}\, (T_{\rm eff}/20000),
\end{split}
\end{equation}
with $v_{\infty}/v_{\rm esc}=1.3$ for 12500 K $\leq T_{\rm eff}\leq 22500\,\rm K$. 

The mass loss in the Wind2 prescription is taken to be
\begin{equation}
\dot{M}={\rm max} (\dot{M}_{\rm NJ},\dot{M}_{\rm R},\dot{M}_{\rm VW},\dot{M}_{\rm WR},\dot{M}_{\rm OB}).
\end{equation}

We ignore the effect of stellar rotation in the calculation, because CE evolution usually occurs when the donor star has already entered the giant phase with slow rotation.

\begin{figure}
 \includegraphics[width=3.0in]{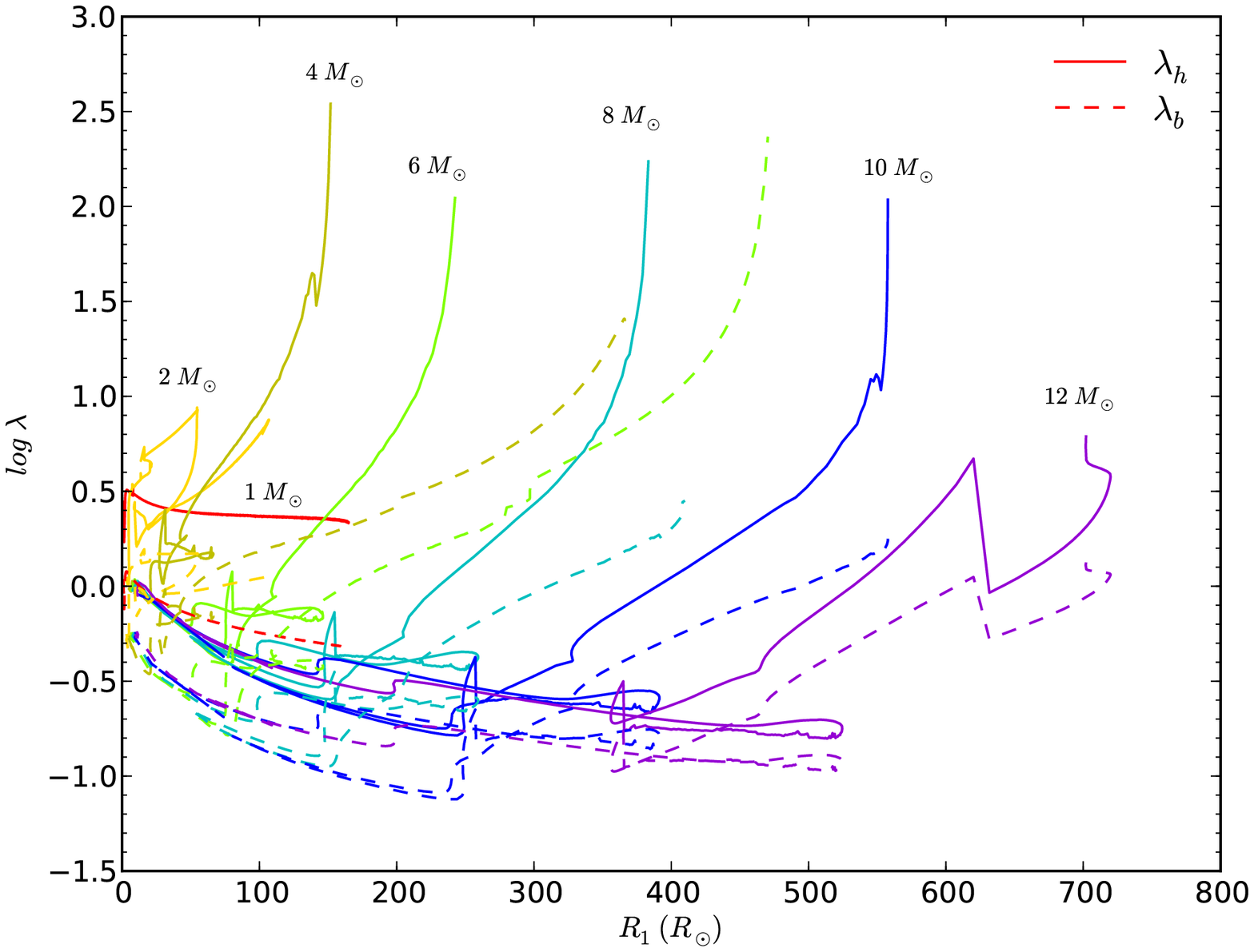}
 \includegraphics[width=3.0in]{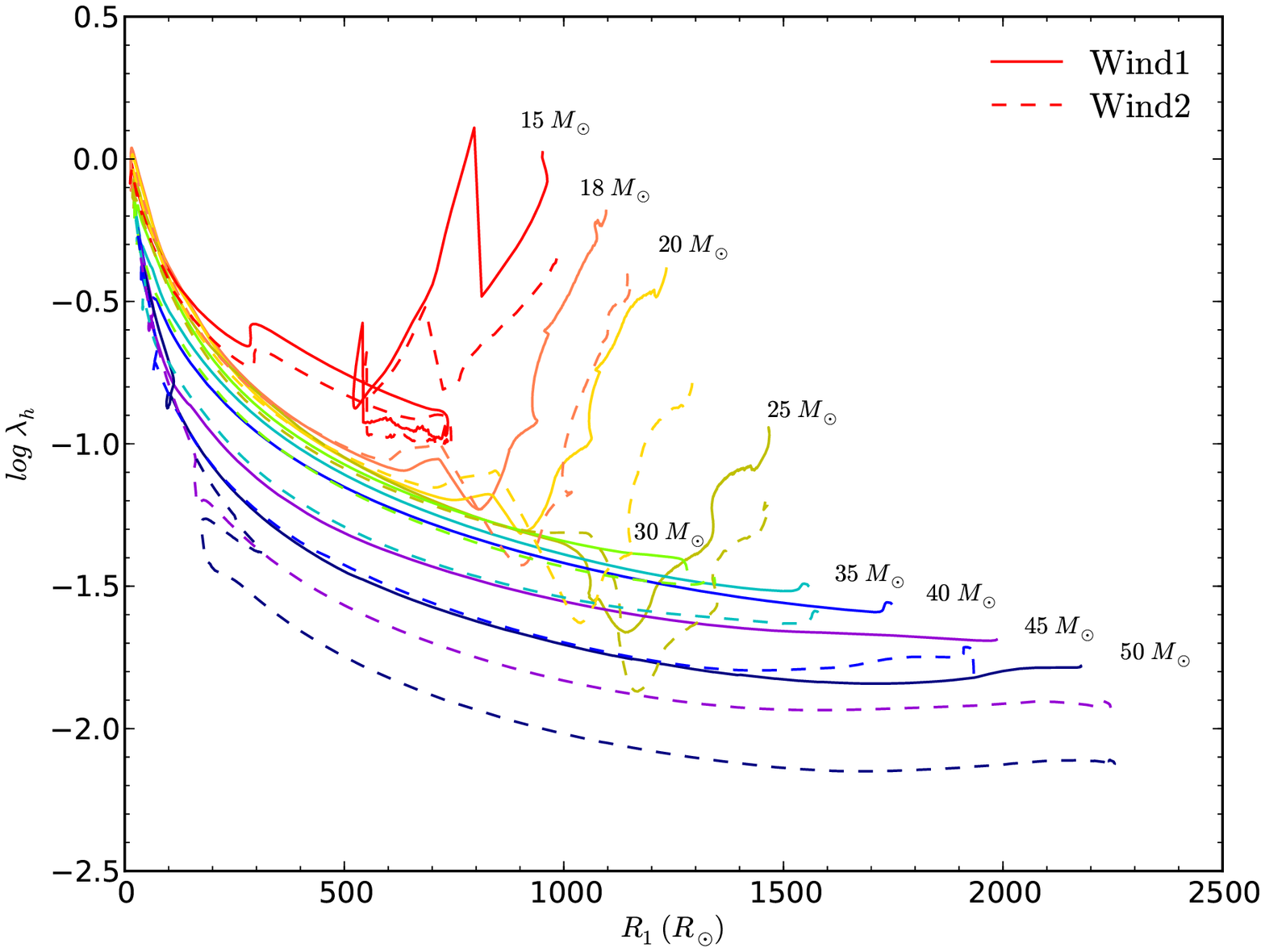}
\caption{Comparison of the values of $\lambda$ for different stars. The left panel the solid and dashed lines represent $\lambda_{\rm h}$ and $\lambda_{\rm b}$ for stars less massive than $15\, M_{\sun}$ with the Wind1 prescription, respectively. The right panel shows $\lambda_{\rm h}$ as a function of the stellar radius for stars more massive than $15\, M_{\sun}$.  The solid and dashed lines represent the results with the Wind1 and Wind2 prescriptions, respectively. \label{figure3}}
\end{figure}

\section{Results and discussion}
We have calculated the values of different $\lambda$s for $1-60\, M_{\sun}$ stars by combining Eqs.~[2]-[5], considering gravitational binding energy alone, total energy, and total energy plus enthalpy in the stellar envelope separately. We denote them as $\lambda_{\rm g}$, $\lambda_{\rm b}$ and $\lambda_{\rm h}$, respectively, that is,
\[-\frac {GM_1 M_{\rm env}}{\lambda_{\rm g} a_{\rm i} r_{\rm L}}=-\int_{M_{\rm core}}^{M_1}\frac{GM(r)}{r}dm,\]
\[-\frac {GM_1 M_{\rm env}}{\lambda_{\rm b} a_{\rm i} r_{\rm L}}=\int_{M_{\rm core}}^{M_1}\left[-\frac{GM(r)}{r}+U\right ] dm,\] and \[-\frac {GM_1 M_{\rm env}}{\lambda_{\rm h} a_{\rm i} r_{\rm L}}=\int_{M_{\rm core}}^{M_1}\left[-\frac{GM(r)}{r}+U+\frac{P}{\rho}\right] dm. \] More massive stars may lose most of their envelope through strong winds so CE evolution is not likely to occur. For each star, we follow its evolution until it ascends the thermally pulsating asymptotic giant branch (TP-AGB) where the star initiates repeating expansions and contractions (for stars $>4\, M_{\odot}$), or the code crushes automatically. The binding energy between the envelope and the maximum compression point in the hydrogen burning shell is calculated once a star evolves to produce such a shell. Here, the maximum compression point is the place with the highest local sonic velocity (i.e., the largest value for $P/\rho$ in the shell).

Figures~\ref{figure1} and ~\ref{figure2} show the evolution of $\lambda$s with respect to the stellar radius $R$ for stars with different masses. The solid and dashed lines represent the results under the Wind1 and Wind2 prescriptions, respectively. The green, blue, and red lines correspond to $\lambda_{\rm g}$, $\lambda_{\rm b}$ and $\lambda_{\rm h}$ respectively in each case. In general they demonstrate similar evolutionary trend, with $\lambda_{\rm b}$ being roughly twice as large as $\lambda_{\rm g}$, which is a natural consequence of the Viral theorem, and $\lambda_{\rm h}$ being several times larger than $\lambda_{\rm b}$.
For stars with mass $\sim 1\, M_{\sun}$ or $>30\, M_{\sun}$, $\lambda$s decrease constantly along the evolutionary tracks, while for stars with mass in between,  $\lambda$s decrease with increasing $R$ at first and then increase when they have ascended the AGB and developed a deep convective envelope (see also \citealt{Podsiadlowski2003}). Most interestingly, for $\sim 3-10\, M_{\sun}$ stars, $\lambda_{\rm h}$ (and $\lambda_{\rm b}$ in some cases) increases drastically in the supergiant phase, and can reach a ``boiling pot" zone (see \citealt{Han1994,Ivanova2011}), where the  binding energy becomes positive before it expands to the reach the maximum radius.

The differences between the solid and dashed lines demonstrate the influence of wind loss on the mass and the compactness of the envelope, especially near the core-envelope boundary. We can see that for stars of mass $\lesssim 15\, M_{\sun}$ (except $1\, M_{\sun}$), or stars of mass $\sim 15-30\, M_{\sun}$ but with radius $\lesssim 500-1000R_\odot$, the $\lambda$-values in the Wind1 case roughly coincide with those in the Wind2 case, reflecting that the two prescriptions are almost the same in such situations. In other cases, the $\lambda$-values in the Wind2 case are usually smaller than in the Wind1 case because of steeper density profile in the envelope \citep[see also][]{Podsiadlowski2003}.

To see how the binding energy changes with stellar mass, we compare $\lambda_{\rm h}$  and $\lambda_{\rm b}$  as a function of $R$ for stars with different masses in Fig.~\ref{figure3}. Generally more massive stars have smaller $\lambda$,  implying that ejection of the envelope is more difficult. This feature is particularly important for the formation of black hole low-mass X-ray binaries \citep[][and references there in]{Justham2006,Wang2016}.

Our calculated binding energy parameters $\lambda_{\rm g}$ and $\lambda_{\rm b}$ evolve in a way in general accord with in previous studies.  However, there are some remarkable differences in specific circumstances. (1) Comparing with \cite{XuLi2010a,XuLi2010b}, we find that for $2-8\, M_{\odot}$ stars the values of $\lambda_{\rm g}$ and $\lambda_{\rm b}$  increase more rapidly with radius during the AGB stage. For example, for a $3\,M_{\odot}$ star, $\lambda_{\rm b} \simeq 10$ at $300\, R_{\odot}$ in \cite{XuLi2010a,XuLi2010b}, but $\simeq 100 $ in our case. (2) Our calculations show that the $\lambda$-values increase with radius at the very end of the evolutionary stages for stars less massive than $\sim 30\, M_{\odot}$ (except for $1\, M_{\odot}$ star), while this upper mass limit becomes lower, i.e., $\sim 20\, M_{\odot}$  in \cite{XuLi2010a,XuLi2010b} and \cite{Podsiadlowski2003}. (3) The $\lambda$-values do not show a significant decline at the very end of the evolution for stars more massive than $\sim 30\, M_{\odot}$ as observed by \cite{Podsiadlowski2003}. 

Finally, similar as \citet{XuLi2010a,XuLi2010b} we perform polynomial fitting for the calculated $\lambda$-values,
\begin{equation}\label{equation7}
   {\rm log}\,  \lambda=a_{0}+a_{1}x+a_{2}x^{2}+a_{3}x^{3}+a_{4}x^{4}+a_{5}x^{5}+a_{6}x^{6},
\end{equation}
where $x=R/R_{\odot}$.
For $2-15\, M_{\sun}$ stars that have ``hook''-like features in $\lambda$, we divide its post-main-sequence evolution into three stages, and fit the $\lambda$-values separately. Stage 1 begins at the exhaustion of central hydrogen and ends when the star starts to shrink (i.e., near the ignition of central He). Stage 2 is the following shrinking phase, and in stage 3 the star expands again, until the end of the evolution.
In Table~\ref{table1}, we list the fitting parameters. We use the coefficient of determination $\mathcal{R}^2$ (i.e., the ratio of the regression sum of squares to the total sum of squares) to evaluate the goodness of fit: $\mathcal{R}^2=1$ corresponds to perfect fit, while $\mathcal{R}^2=0$ indicates that the equation does not fit the data at all. In our fitting results, the values of $\mathcal{R}^2$ in all the cases are above 0.95.

\renewcommand{\tablename}{\normalsize Table}
\begin{tiny}
\begin{longtable}{ccccccccccc}

\caption{\normalsize Fitting parameters for $\lambda$. \label{table1}}\\
\hline
\hline
  Mass $(M_{\sun})$  &  Wind loss   & stage &   $\lambda$   & $a_{0}$ & $a_{1}$   &   $a_{2}$   & $a_{3}$ & $a_{4}$ & $a_{5}$  & $a_{6}$ \\
  \hline
  \endfirsthead
  \caption[]{\normalsize Fitting parameters for $\lambda$ (continued).}\\
\hline
\hline
  Mass $(M_{\sun})$  &  Wind loss   & stage &  $\lambda$   & $a_{0}$ & $a_{1}$   &   $a_{2}$   & $a_{3}$ & $a_{4}$ & $a_{5}$  & $a_{6}$ \\
  \hline
  \endhead
  \hline
  \endfoot
  \hline\hline\endlastfoot
 1	&	Wind1	&		&	$\lambda_{\rm h}$	&	 0.439377368	&	0.006334748	&	-4.60E-04	&	 1.05E-05	&	 -1.11E-07	&	5.51E-10	&	 -1.05E-12	\\
	&		&		&	$\lambda_{\rm b}$	&	 0.047486107	&	-0.001813814	&	-0.000202148	 &	5.51E-06	&	 -6.12E-08	&	3.13E-10	&	 -6.06E-13	\\
	&		&		&	$\lambda_{\rm g}$	&	 -0.229504942	&	-0.006642966	&	-7.90E-05	 &	3.36E-06	&	 -4.09E-08	&	2.18E-10	&	 -4.30E-13	\\
1	&	Wind2	&		&	$\lambda_{\rm h}$	&	 0.446551051	&	0.004283965	&	-0.000362559	&	 7.71E-06	&	 -7.78E-08	&	3.68E-10	&	 -6.63E-13	\\
	&		&		&	$\lambda_{\rm b}$	&	 0.051723323	&	-0.003092511	&	-0.000145916	 &	3.91E-06	&	 -4.21E-08	&	2.05E-10	&	 -3.78E-13	\\
	&		&		&	$\lambda_{\rm g}$	&	 -0.226240947	&	-0.007668935	&	-3.70E-05	 &	2.17E-06	&	 -2.67E-08	&	1.37E-10	&	 -2.60E-13	\\
\hline																					
2	&	Wind1	&	1	&	$\lambda_{\rm h}$	&	0.458299689	&	-0.066491722	&	0.012809574	&	-0.00079697	&	2.35E-05	&	-3.33E-07	&	1.83E-09	\\
	&		&		&	$\lambda_{\rm b}$	&	-0.055628327	&	-0.020278881	&	0.006650167	&	-0.000473348	&	1.49E-05	&	-2.18E-07	&	1.23E-09	\\
	&		&		&	$\lambda_{\rm g}$	&	-0.365205393	&	-0.008048466	&	0.004718683	&	-0.000366995	&	1.19E-05	&	-1.79E-07	&	1.02E-09	\\
	&		&	2	&	$\lambda_{\rm h}$	&	1.503848844	&	-0.307827791	&	0.033926633	&	-0.001915136	&	5.72E-05	&	-8.51E-07	&	4.97E-09	\\
	&		&		&	$\lambda_{\rm b}$	&	0.078894909	&	-0.017355261	&	0.001880501	&	-0.000145636	&	5.81E-06	&	-1.06E-07	&	7.16E-10	\\
	&		&		&	$\lambda_{\rm g}$	&	-0.318594414	&	0.01353053	&	-0.001715174	&	6.52E-05	&	-7.02E-07	&	-6.60E-09	&	1.24E-10	\\
	&		&	3	&	$\lambda_{\rm h}$	&	0.80094725	&	-0.088162823	&	0.005190546	&	-0.000134579	&	1.77E-06	&	-1.14E-08	&	2.90E-11	\\
	&		&		&	$\lambda_{\rm b}$	&	0.042516824	&	-0.017189807	&	0.000736229	&	-1.26E-05	&	1.02E-07	&	-3.85E-10	&	5.44E-13	\\
	&		&		&	$\lambda_{\rm g}$	&	-0.278858406	&	-0.011175318	&	0.000408553	&	-6.72E-06	&	5.31E-08	&	-1.95E-10	&	2.66E-13	\\
\hline																					
3	&	Wind1	&	1	&	$\lambda_{\rm h}$	&	1.32158616	&	-0.511107713	&	0.069798235	&	-0.004234453	&	0.000129527	&	-1.96E-06	&	1.16E-08	\\
	&		&		&	$\lambda_{\rm b}$	&	0.553601499	&	-0.322402445	&	0.041379969	&	-0.002281134	&	6.15E-05	&	-7.87E-07	&	3.71E-09	\\
	&		&		&	$\lambda_{\rm g}$	&	0.147496542	&	-0.255352018	&	0.031009238	&	-0.001567046	&	3.67E-05	&	-3.63E-07	&	8.52E-10	\\
	&		&	2	&	$\lambda_{\rm h}$	&	-34.47232944	&	8.284562942	&	-0.792027265	&	0.039213067	&	-0.001063886	&	1.50E-05	&	-8.66E-08	\\
	&		&		&	$\lambda_{\rm b}$	&	-21.26420026	&	4.998641287	&	-0.47435718	&	0.023385099	&	-0.000633095	&	8.94E-06	&	-5.15E-08	\\
	&		&		&	$\lambda_{\rm g}$	&	-17.03794141	&	3.912297317	&	-0.369651011	&	0.018176049	&	-0.000491475	&	6.94E-06	&	-4.00E-08	\\
	&		&	3	&	$\lambda_{\rm h}$	&	1.761358786	&	-0.210912284	&	0.010365416	&	-0.000222355	&	2.40E-06	&	-1.29E-08	&	2.73E-11	\\
	&		&		&	$\lambda_{\rm b}$	&	-0.304912365	&	0.016425641	&	-0.000214159	&	9.87E-07	&	1.30E-09	&	-2.05E-11	&	3.98E-14	\\
	&		&		&	$\lambda_{\rm g}$	&	-0.556787574	&	0.013693598	&	-0.000219799	&	1.53E-06	&	-4.85E-09	&	5.74E-12	&	5.16E-17	\\
\hline																					
																					
4	&	Wind1	&	1	&	$\lambda_{\rm h}$	&	1.060028729	&	-0.259126108	&	0.021464041	&	-0.000824166	&	1.64E-05	&	-1.63E-07	&	6.48E-10	\\
	&		&		&	$\lambda_{\rm b}$	&	0.396069457	&	-0.160359901	&	0.010528477	&	-0.000277383	&	2.93E-06	&	-4.06E-09	&	-7.92E-11	\\
	&		&		&	$\lambda_{\rm g}$	&	0.030959831	&	-0.128292612	&	0.006840813	&	-9.09E-05	&	-1.69E-06	&	5.09E-08	&	-3.31E-10	\\
	&		&	2	&	$\lambda_{\rm h}$	&	-158.2144537	&	21.3665151	&	-1.175629306	&	0.033828851	&	-0.000537603	&	4.48E-06	&	-1.53E-08	\\
	&		&		&	$\lambda_{\rm b}$	&	-113.5126224	&	15.15018357	&	-0.827530643	&	0.023676471	&	-0.000374536	&	3.11E-06	&	-1.06E-08	\\
	&		&		&	$\lambda_{\rm g}$	&	-95.30490909	&	12.61279533	&	-0.686025008	&	0.019562306	&	-0.000308629	&	2.56E-06	&	-8.69E-09	\\
	&		&	3	&	$\lambda_{\rm h}$	&	6.21658444	&	-0.610564473	&	0.022771632	&	-0.000409896	&	3.88E-06	&	-1.86E-08	&	3.55E-11	\\
	&		&		&	$\lambda_{\rm b}$	&	-0.865298169	&	0.023395321	&	-0.000167028	&	2.40E-07	&	2.51E-09	&	-1.07E-11	&	1.22E-14	\\
	&		&		&	$\lambda_{\rm g}$	&	-1.318602631	&	0.035303846	&	-0.000452023	&	2.89E-06	&	-9.71E-09	&	1.63E-11	&	-1.08E-14	\\

\hline																					
6	&	Wind1	&	1	&	$\lambda_{\rm h}$	&	 -0.080220619	&	0.018717693	&	-0.001599316	 &	4.43E-05	&	 -5.53E-07	&	3.21E-09	&	 -7.08E-12	\\
	&		&		&	$\lambda_{\rm b}$	&	 -0.181116753	&	-0.007930443	&	 -0.000669357	&	2.93E-05	&	 -4.24E-07	&	 2.63E-09	&	-6.00E-12	\\
	&		&		&	$\lambda_{\rm g}$	&	 -0.333386602	&	-0.022017403	&	 -0.00015568	&	2.03E-05	&	 -3.38E-07	&	 2.21E-09	&	-5.17E-12	\\
	&		&	2	&	$\lambda_{\rm h}$	&	 73.5586189	&	-4.777609245	&	0.124527396	&	 -0.001679558	&	 1.24E-05	&	-4.79E-08	&	 7.56E-11	\\
	&		&		&	$\lambda_{\rm b}$	&	 77.90321325	&	-4.981465802	&	0.126967686	&	 -0.001673014	&	 1.21E-05	&	-4.56E-08	&	 7.05E-11	\\
	&		&		&	$\lambda_{\rm g}$	&	 77.71103944	&	-4.947175949	&	0.124917607	&	 -0.001629579	&	 1.17E-05	&	-4.36E-08	&	 6.66E-11	\\
	&		&	3	&	$\lambda_{\rm h}$	&	 6.445886556	&	-0.314136789	&	0.005644353	&	 -5.18E-05	&	 2.67E-07	&	-7.40E-10	&	 8.60E-13	\\
	&		&		&	$\lambda_{\rm b}$	&	 1.958799141	&	-0.11166705	&	0.001646683	&	 -1.08E-05	&	 3.63E-08	 &	-6.01E-11	&	 3.92E-14	\\
	&		&		&	$\lambda_{\rm g}$	&	 0.571516609	&	-0.06864172	&	0.001021307	&	 -6.51E-06	&	 2.08E-08	 &	-3.27E-11	&	 2.02E-14	\\
\hline																					
8	&	Wind1	&	1	&	$\lambda_{\rm h}$	&	 -0.035692021	&	0.005572222	&	-0.000485211	 &	8.19E-06	&	 -5.94E-08	&	1.97E-10	&	 -2.46E-13	\\
	&		&		&	$\lambda_{\rm b}$	&	 -0.167246715	&	-0.00963629	&	-0.000147715	 &	4.83E-06	&	 -4.17E-08	&	1.50E-10	&	 -1.95E-13	\\
	&		&		&	$\lambda_{\rm g}$	&	 -0.332644809	&	-0.019069101	&	6.84E-05	 &	2.52E-06	&	 -2.87E-08	&	1.12E-10	&	 -1.52E-13	\\
	&		&	2	&	$\lambda_{\rm h}$	&	 -2.365068424	&	0.124227402	&	-0.002735286	 &	2.83E-05	&	 -1.50E-07	&	3.93E-10	&	 -4.04E-13	\\
	&		&		&	$\lambda_{\rm b}$	&	 -4.533738779	&	0.212107478	&	-0.004366164	 &	4.29E-05	&	 -2.17E-07	&	5.44E-10	&	 -5.37E-13	\\
	&		&		&	$\lambda_{\rm g}$	&	 -5.641282203	&	0.250488572	&	-0.005065902	 &	4.90E-05	&	 -2.44E-07	&	6.02E-10	&	 -5.86E-13	\\
	&		&	3	&	$\lambda_{\rm h}$	&	 0.414442304	&	-0.025433442	&	0.000380787	&	 -3.75E-06	&	 2.07E-08	&	-5.39E-11	&	 5.25E-14	\\
	&		&		&	$\lambda_{\rm b}$	&	0.288646981     &  	-0.026900813	&0.000212298	&-7.41E-07	&1.58E-09	&-2.06E-12	&1.25E-15  \\
	&		&		&	$\lambda_{\rm g}$	&	0.119652903   &    -0.031712235	& 0.000260052	&-8.94E-07	&1.57E-09	&-1.39E-12	&4.91E-16    \\
\hline																					
10	&	Wind1	&	1	&	$\lambda_{\rm h}$	&	 -0.018057871	&	-0.000377557	&	 -0.000140871	&	1.79E-06	 &	 -9.01E-09	&	 2.03E-11	&	-1.70E-14	\\
	&		&		&	$\lambda_{\rm b}$	&	 -0.190593228	&	-0.008930875	&	8.17E-06	 &	6.41E-07	&	 -4.49E-09	&	1.14E-11	&	 -1.01E-14	\\
	&		&		&	$\lambda_{\rm g}$	&	 -0.380447742	&	-0.015346001	&	 0.000118835	&	-2.39E-07	&	 -8.97E-10	&	 4.20E-12	&	-4.47E-15	\\
	&		&	2	&	$\lambda_{\rm h}$	&	 -3.620269125	&	0.112281354	&	-0.001478372	 &	9.22E-06	&	 -2.96E-08	&	4.74E-11	&	 -2.99E-14	\\
	&		&		&	$\lambda_{\rm b}$	&	 -4.030432414	&	0.107086292	&	-0.001308216	 &	7.39E-06	&	 -2.07E-08	&	2.77E-11	&	 -1.36E-14	\\
	&		&		&	$\lambda_{\rm g}$	&	 -4.196331979	&	0.09768744	&	-0.001127814	 &	5.80E-06	&	 -1.38E-08	&	1.33E-11	&	 -2.25E-15	\\
	&		&	3	&	$\lambda_{\rm h}$	&	 0.870255685	&	-0.035282331	&	0.000383239	&	 -2.32E-06	&	 7.54E-09	&	-1.20E-11	&	 7.31E-15	\\
	&		&		&	$\lambda_{\rm b}$	&	 -1.908657748	&	0.037327062	&	-0.000439988	 &	2.22E-06	&	 -5.37E-09	&	6.26E-12	&	 -2.83E-15	\\
	&		&		&	$\lambda_{\rm g}$	&	 -2.592139763	&	0.048484157	&	-0.000573382	 &	2.98E-06	&	 -7.57E-09	&	9.32E-12	&	 -4.48E-15	\\
\hline																					
12	&	Wind1	&	1	&	$\lambda_{\rm h}$	&	 0.003046098	&	-0.002954184	&	-4.77E-05	&	 5.48E-07	&	 -2.16E-09	&	3.69E-12	&	 -2.31E-15	\\
	&		&		&	$\lambda_{\rm b}$	&	 -0.179897401	&	-0.007866116	&	1.32E-05	 &	2.14E-07	&	 -1.20E-09	&	2.31E-12	&	 -1.53E-15	\\
	&		&		&	$\lambda_{\rm g}$	&	 -0.382469592	&	-0.01241696	&	6.94E-05	&	 -1.11E-07	&	 -2.31E-10	 &	8.69E-13	&	 -6.91E-16	\\
	&		&	2	&	$\lambda_{\rm h}$	&	 -3527.277445	&	49.22217981	&	-0.28437734	&	 0.000870754	&	 -1.49E-06	&	1.35E-09	&	 -5.09E-13	\\
	&		&		&	$\lambda_{\rm b}$	&	 -4066.073674	&	56.0114003	&	-0.319954463	 &	0.000969959	&	 -1.65E-06	&	1.48E-09	&	 -5.54E-13	\\
	&		&		&	$\lambda_{\rm g}$	&	 -4223.887667	&	57.81457007	&	-0.328432188	 &	0.000990866	&	 -1.67E-06	&	1.50E-09	&	 -5.60E-13	\\
	&		&	3	&	$\lambda_{\rm h}$	&	 -1123.405701	&	13.39173279	&	-0.065527774	 &	0.000168289	&	 -2.39E-07	&	1.79E-10	&	 -5.48E-14	\\
	&		&		&	$\lambda_{\rm b}$	&	 -650.9073393	&	7.754985137	&	-0.037961302	 &	9.75E-05	&	 -1.39E-07	&	1.03E-10	&	 -3.17E-14	\\
	&		&		&	$\lambda_{\rm g}$	&	 -408.154727	&	4.861180309	&	-0.02384045	&	 6.14E-05	&	 -8.74E-08	 &	6.54E-11	&	 -2.01E-14	\\
\hline																					
15	&	Wind1	&	1	&	$\lambda_{\rm h}$	&	 -0.028115873	&	-0.002865229	&	-2.06E-05	 &	1.88E-07	&	 -5.34E-10	&	6.45E-13	&	 -2.85E-16	\\
	&		&		&	$\lambda_{\rm b}$	&	 -0.189058198	&	-0.006845831	&	1.38E-05	 &	6.09E-08	&	 -2.87E-10	&	4.01E-13	&	 -1.89E-16	\\
	&		&		&	$\lambda_{\rm g}$	&	 -0.384916072	&	-0.0111889	&	5.15E-05	&	 -8.88E-08	&	 2.15E-11	 &	8.41E-14	&	 -6.07E-17	\\
	&		&	2	&	$\lambda_{\rm h}$	&	 -43390.80866	&	415.5416274	&	-1.652720808	 &	0.003494469	&	 -4.14E-06	&	2.61E-09	&	 -6.84E-13	\\
	&		&		&	$\lambda_{\rm b}$	&	 -38341.9421	&	366.9224492	&	-1.458469908	&	 0.00308221	&	 -3.65E-06	&	2.30E-09	&	 -6.03E-13	\\
	&		&		&	$\lambda_{\rm g}$	&	 -30256.13593	&	289.1537559	&	-1.14796062	&	 0.002423327	&	 -2.87E-06	&	1.81E-09	&	 -4.72E-13	\\
	&		&	3	&	$\lambda_{\rm h}$	&	 401.3761414	&	-3.338142889	&	0.011206254	&	 -1.93E-05	&	 1.77E-08	&	-7.96E-12	&	 1.27E-15	\\
	&		&		&	$\lambda_{\rm b}$	&	 511.5727996	&	-4.483039299	&	0.016065894	&	 -3.01E-05	&	 3.11E-08	&	-1.67E-11	&	 3.61E-15	\\
	&		&		&	$\lambda_{\rm g}$	&	 756.3447803	&	-6.901553717	&	0.025936812	&	 -5.15E-05	&	 5.69E-08	&	-3.32E-11	&	 7.97E-15	\\
\hline																					
18	&	Wind1	&		&	$\lambda_{\rm h}$	&	 -0.011555316	&	-0.00112965	&	-3.41E-05	&	 1.73E-07	&	 -3.37E-10	&	2.89E-13	&	 -9.07E-17	\\
	&		&		&	$\lambda_{\rm b}$	&	 -1.93E-01	&	-0.005828186	&	1.91E-06	&	 5.64E-08	&	 -1.57E-10	 &	1.58E-13	&	 -5.41E-17	\\
	&		&		&	$\lambda_{\rm g}$	&	 -4.28E-01	&	-0.010179044	&	3.42E-05	&	 -4.69E-08	&	 4.25E-12	 &	3.87E-14	&	 -2.07E-17	\\
	&	Wind2	&		&	$\lambda_{\rm h}$	&	 -0.085914616	&	0.000704026	&	-4.89E-05	&	 2.11E-07	&	 -3.67E-10	&	2.83E-13	&	 -7.97E-17	\\
	&		&		&	$\lambda_{\rm b}$	&	 -0.218521071	&	-0.004292671	&	-1.54E-05	 &	1.15E-07	&	 -2.30E-10	&	1.90E-13	&	 -5.53E-17	\\
	&		&		&	$\lambda_{\rm g}$	&	 -0.40399817	&	-0.009339995	&	1.81E-05	&	 1.74E-08	&	 -9.16E-11	&	9.45E-14	&	 -3.03E-17	\\
\hline																					
20	&	Wind1	&		&	$\lambda_{\rm h}$	&	 3.33E-02	&	-2.77E-03	&	-1.69E-05	&	 9.48E-08	&	 -1.77E-10	 &	1.40E-13	&	 -4.02E-17	\\
	&		&		&	$\lambda_{\rm b}$	&	 -0.236292531	&	-4.92E-03	&	-6.45E-07	&	 4.58E-08	&	 -1.07E-10	 &	9.55E-14	&	 -2.93E-17	\\
	&		&		&	$\lambda_{\rm g}$	&	 -0.57266423	&	-0.006998902	&	1.41E-05	&	 1.46E-09	&	 -4.40E-11	&	5.30E-14	&	 -1.86E-17	\\
	&	Wind2	&		&	$\lambda_{\rm h}$	&	 -0.084425289	&	-0.000248147	&	-3.41E-05	 &	1.34E-07	&	 -2.05E-10	&	1.36E-13	&	 -3.30E-17	\\
	&		&		&	$\lambda_{\rm b}$	&	 -0.352025741	&	-0.001677298	&	-2.49E-05	 &	1.09E-07	&	 -1.72E-10	&	1.16E-13	&	 -2.83E-17	\\
	&		&		&	$\lambda_{\rm g}$	&	 -0.684632138	&	-0.00347904	&	-1.36E-05	&	 7.81E-08	&	 -1.30E-10	 &	9.01E-14	&	 -2.21E-17	\\
\hline																					
25	&	Wind1	&		&	$\lambda_{\rm h}$	&	 -0.079805635	&	-0.001539374	&	-1.62E-05	 &	6.42E-08	&	 -9.17E-11	&	5.65E-14	&	 -1.26E-17	\\
	&		&		&	$\lambda_{\rm b}$	&	 -0.217137659	&	-0.005291635	&	4.59E-06	 &	1.57E-08	&	 -3.82E-11	&	2.89E-14	&	 -7.22E-18	\\
	&		&		&	$\lambda_{\rm g}$	&	 -0.445989682	&	-0.009416417	&	2.67E-05	 &	-3.53E-08	&	 1.83E-11	&	-7.71E-16	&	 -1.27E-18	\\
	&	Wind2	&		&	$\lambda_{\rm h}$	&	 -1.21E-01	&	-7.73E-04	&	-2.46E-05	&	 8.75E-08	&	 -1.17E-10	 &	6.73E-14	&	 -1.40E-17	\\
	&		&		&	$\lambda_{\rm b}$	&	 -2.26E-01	&	-4.80E-03	&	-2.55E-06	&	 3.71E-08	&	-6.20E-11	 &	 3.94E-14	&	 -8.66E-18	\\
	&		&		&	$\lambda_{\rm g}$	&	 -4.07E-01	&	-9.64E-03	&	2.31E-05	&	 -2.12E-08	&	1.77E-12	 &	 6.23E-15	&	 -2.05E-18	\\
\hline																					
30	&	Wind1	&		&	$\lambda_{\rm h}$	&	 -0.068721052	&	-0.004439298	&	7.66E-06	 &	-5.86E-09	&	 -9.70E-13	&	3.75E-15	&	 -1.41E-18	\\
	&		&		&	$\lambda_{\rm b}$	&	 -0.168597281	&	-0.008185857	&	2.91E-05	 &	-5.99E-08	&	 6.65E-11	&	-3.71E-14	&	 8.18E-18	\\
	&		&		&	$\lambda_{\rm g}$	&	 -0.360619104	&	-0.013289713	&	5.80E-05	 &	-1.33E-07	&	 1.57E-10	&	-9.18E-14	&	 2.10E-17	\\
	&	Wind2	&		&	$\lambda_{\rm h}$	&	 -2.64E-02	&	-8.04E-03	&	3.08E-05	&	 -6.72E-08	&	 7.72E-11	 &	-4.41E-14	&	 9.86E-18	\\
	&		&		&	$\lambda_{\rm b}$	&	 -1.16E-01	&	-1.10E-02	&	4.60E-05	&	 -1.02E-07	&	1.18E-10	 &	 -6.72E-14	&	 1.50E-17	\\
	&		&		&	$\lambda_{\rm g}$	&	 -2.79E-01	&	-1.57E-02	&	7.00E-05	&	 -1.58E-07	&	1.83E-10	 &	 -1.04E-13	&	 2.30E-17	\\
\hline																					
35	&	Wind1	&		&	$\lambda_{\rm h}$	&	 -0.086834231	&	-0.005683759	&	1.61E-05	 &	-2.79E-08	&	 2.67E-11	&	-1.29E-14	&	 2.48E-18	\\
	&		&		&	$\lambda_{\rm b}$	&	 -0.178466739	&	-0.008221884	&	2.75E-05	 &	-5.09E-08	&	 4.98E-11	&	-2.42E-14	&	 4.64E-18	\\
	&		&		&	$\lambda_{\rm g}$	&	 -0.368625486	&	-0.012306112	&	4.58E-05	 &	-8.80E-08	&	 8.71E-11	&	-4.25E-14	&	 8.12E-18	\\
	&	Wind2	&		&	$\lambda_{\rm h}$	&	 4.10E-03	&	-1.07E-02	&	4.09E-05	&	 -8.02E-08	&	 8.04E-11	 &	-3.95E-14	&	 7.55E-18	\\
	&		&		&	$\lambda_{\rm b}$	&	 -0.082606521	&	-1.23E-02	&	4.72E-05	&	 -9.18E-08	&	 9.13E-11	 &	-4.46E-14	&	 8.48E-18	\\
	&		&		&	$\lambda_{\rm g}$	&	 -0.244193403	&	-0.01566844	&	6.06E-05	&	 -1.17E-07	&	 1.15E-10	 &	-5.57E-14	&	 1.05E-17	\\
\hline																					
40	&	Wind1	&		&	$\lambda_{\rm h}$	&	 -0.113572735	&	-0.006236406	&	1.80E-05	 &	-2.96E-08	&	 2.58E-11	&	-1.13E-14	&	 1.94E-18	\\
	&		&		&	$\lambda_{\rm b}$	&	 -0.183810999	&	-0.008309639	&	2.59E-05	 &	-4.35E-08	&	 3.81E-11	&	-1.66E-14	&	 2.83E-18	\\
	&		&		&	$\lambda_{\rm g}$	&	 -0.349966999	&	-0.012034168	&	4.03E-05	 &	-6.88E-08	&	 6.05E-11	&	-2.62E-14	&	 4.45E-18	\\
	&	Wind2	&		&	$\lambda_{\rm h}$	&	 1.44E-03	&	-1.08E-02	&	3.30E-05	&	 -5.04E-08	&	 3.92E-11	 &	-1.49E-14	&	 2.21E-18	\\
	&		&		&	$\lambda_{\rm b}$	&	 -1.10E-01	&	-1.12E-02	&	3.43E-05	&	 -5.22E-08	&	4.05E-11	 &	 -1.55E-14	&	 2.30E-18	\\
	&		&		&	$\lambda_{\rm g}$	&	 -3.19E-01	&	-1.28E-02	&	3.90E-05	&	 -5.91E-08	&	4.58E-11	 &	 -1.75E-14	&	 2.62E-18	\\
\hline																					
45	&	Wind1	&		&	$\lambda_{\rm h}$	&	 -0.09033098	&	-0.0078733	&	2.16E-05	&	 -3.12E-08	&	 2.35E-11	&	-8.79E-15	&	 1.29E-18	\\
	&		&		&	$\lambda_{\rm b}$	&	 -0.165033235	&	-0.00897789	&	2.51E-05	&	 -3.64E-08	&	 2.74E-11	 &	-1.03E-14	&	 1.50E-18	\\
	&		&		&	$\lambda_{\rm g}$	&	 -0.334563352	&	-0.011618318	&	3.36E-05	 &	-4.94E-08	&	 3.75E-11	&	-1.40E-14	&	 2.06E-18	\\
	&	Wind2	&		&	$\lambda_{\rm h}$	&	 -5.84E-02	&	-9.31E-03	&	2.27E-05	&	 -2.86E-08	&	 1.87E-11	 &	-6.06E-15	&	 7.70E-19	\\
	&		&		&	$\lambda_{\rm b}$	&	 -1.76E-01	&	-9.24E-03	&	2.22E-05	&	 -2.76E-08	&	1.79E-11	 &	 -5.77E-15	&	 7.29E-19	\\
	&		&		&	$\lambda_{\rm g}$	&	 -4.28E-01	&	-9.41E-03	&	2.14E-05	&	 -2.57E-08	&	1.62E-11	 &	 -5.09E-15	&	 6.30E-19	\\
\hline																					
50	&	Wind1	&		&	$\lambda_{\rm h}$	&	 -0.077847412	&	-0.008264901	&	2.00E-05	 &	-2.52E-08	&	 1.65E-11	&	-5.39E-15	&	 6.94E-19	\\
	&		&		&	$\lambda_{\rm b}$	&	 -0.173570298	&	-0.008589709	&	2.07E-05	 &	-2.60E-08	&	 1.71E-11	&	-5.58E-15	&	 7.20E-19	\\
	&		&		&	$\lambda_{\rm g}$	&	 -0.387442471	&	-0.009858782	&	2.40E-05	 &	-3.02E-08	&	 1.99E-11	&	-6.55E-15	&	 8.51E-19	\\
	&	Wind2	&		&	$\lambda_{\rm h}$	&	 -7.61E-01	&	-3.18E-03	&	4.45E-06	&	 -4.66E-09	&	 3.00E-12	 &	-9.92E-16	&	 1.29E-19	\\
	&		&		&	$\lambda_{\rm b}$	&	 -8.68E-01	&	-2.98E-03	&	3.68E-06	&	 -3.56E-09	&	2.26E-12	 &	 -7.58E-16	&	 1.00E-19	\\
	&		&		&	$\lambda_{\rm g}$	&	 -1.18E+00	&	-2.19E-03	&	5.71E-07	&	 8.88E-10	&	-7.72E-13	 &	 2.32E-16	&	 -2.38E-20	\\
\hline																					
60	&	Wind1	&		&	$\lambda_{\rm h}$	&	 -0.137506172	&	-0.008191471	&	2.72E-05	 &	-5.69E-08	&	 6.76E-11	&	-4.18E-14	&	 1.04E-17	\\
	&		&		&	$\lambda_{\rm b}$	&	 -0.232113077	&	-0.008216679	&	2.75E-05	 &	-5.78E-08	&	 6.90E-11	&	-4.28E-14	&	 1.07E-17	\\
	&		&		&	$\lambda_{\rm g}$	&	 -0.475332579	&	-0.008678247	&	3.00E-05	 &	-6.42E-08	&	 7.76E-11	&	-4.86E-14	&	 1.22E-17	\\
	&	Wind2	&		&	$\lambda_{\rm h}$	&	 -0.777181069	&	-0.002255006	&	-2.51E-06	 &	9.08E-09	&	 -8.60E-12	&	3.46E-15	&	 -5.13E-19	\\
	&		&		&	$\lambda_{\rm b}$	&	 -0.843092017	&	-0.002433256	&	-2.10E-06	 &	8.63E-09	&	 -8.34E-12	&	3.39E-15	&	 -5.04E-19	\\
	&		&		&	$\lambda_{\rm g}$	&	 -1.055314774	&	-0.00279245	&	-2.09E-06	&	 9.46E-09	&	 -9.29E-12	 &	3.80E-15	&	 -5.66E-19	\\
\end{longtable}
\end{tiny}

\section{Conclusions}
The binding energy parameter $\lambda$ is a key parameter in the formation and evolution of close binary systems. This work is an updated version of \cite{XuLi2010a, XuLi2010b}, with more self-consistent treatments in stellar modeling. The main results are summarized as follows.

1. The $\lambda$-values vary when a star evolves and strongly depends on the star's initial mass. It generally decreases with the increasing stellar radius, but rises at the very end of the evolution for stars less massive than $\sim 30\, M_{\sun}$.

2. More massive stars tend to have smaller $\lambda$. For massive stars ($\gtrsim 15\, M_{\sun}$) the $\lambda$-values are substantially influenced by the wind mass loss.

3. Generally, $\lambda_{\rm h}$ is several times larger than $\lambda_{\rm b}$ and $\lambda_{\rm g}$, which can assist the ejection of the CE. For stars in the mass range of $\sim 3-10\, M_{\sun}$, the $\lambda_{\rm h}$-values can be very large ($> 100$) and even negative before the star reaches its maximum size.

4. Our fitting formulae for $\lambda$s can serve as useful input parameters in population synthesis investigations.

\begin{acknowledgements}
We are grateful to an anonymous referee for helpful comments. This work was funded by the Natural Science Foundation of China under grant numbers 11133001 and 11333004, and the Strategic Priority Research Program of CAS (under grant number XDB09000000).

\end{acknowledgements}


\end{document}